\PassOptionsToPackage{unicode}{hyperref}
\PassOptionsToPackage{hyphens}{url}
\PassOptionsToPackage{dvipsnames,svgnames,x11names}{xcolor}

\documentclass{article}
\DeclareFontFamilySubstitution{TS1}{aer}{cmr}

\usepackage[T1]{fontenc}
\usepackage[utf8]{inputenc}
\usepackage[english]{babel}
\usepackage{geometry}
\usepackage{standalone}
\usepackage{graphicx}
\usepackage[dvipsnames,svgnames,x11names]{xcolor}
\usepackage{mathtools,bm,amssymb}
\usepackage{enumitem}
\usepackage{booktabs}
\usepackage{natbib}
\usepackage{tikz}
\usepackage[strict=true]{csquotes}

\usepackage{setspace}
\usepackage{Sweave}
\usepackage{framed}

\graphicspath{{./figs/}{.}}
\newcommand{\E}{\operatorname{E}}
\newcommand{\pr}{\operatorname{P}}
\newcommand{\PE}{\mathbb{P}}
\newcommand{\Pn}{\mathbb{P}_n}

\newcommand{\CondInd}[3]{#1 \perp\kern-5pt \perp #2 \mid #3}
\newcommand{\Ind}[2]{#1 \perp\kern-5pt \perp #2}

\renewcommand{\subset}{\subseteq}

\newcommand{\independenT}[2]{\mathrel{\setbox0\hbox{$#1#2$}\copy0\kern-\wd0\mkern4mu\box0}}
\newcommand{\indep}{\protect\mathpalette{\protect\independenT}{\perp}}
\newcommand{\esty}[1]{{#1}_{Y\mid T^{\ast}}}
\newcommand*{\estt}[2][]{{#2}_{T^{\ast}#1}}
\newcommand{\potential}[2]{{#1}{}^{(#2)}}
\newcommand{\IT}[1]{I(T^{#1}\geq\tau)}

\begin{document}

% \title{ A framework for joint assessment of treatment effect on occurrence of a terminal event and a score existing only in the absence of the terminal event}

\title{\bf A framework for joint assessment of a terminal event and a score existing only in the absence of the terminal event}
\author{Klaus Kähler Holst${}^1$,
  Andreas Nordland${}^1$,
  Julie Funch Furberg${}^1$, \\
  Lars Holm Damgaard${}^1$,
  Christian Bressen Pipper${}^{1,2}$}
\date{${}^1$Novo Nordisk, Søborg, Denmark \\
  ${}^2$Department of Public Health, Epidemiology, Biostatistics and
  Biodemography, University of Southern Denmark, Odense, Denmark
  \\[2ex]%
  \today}
% \date{\today}
\maketitle

\bigskip

\begin{abstract}
\noindent 
Analysis of data from randomized controlled trials in vulnerable populations requires special
attention when assessing treatment effect by a score measuring, e.g.,  disease stage or activity together with onset of prevalent terminal events. In reality, it is impossible to disentangle a disease score from the terminal event, since the score is not clinically meaningful after this event. In this work, we propose to assess treatment interventions simultaneously on the terminal event and the disease score in the absence of a terminal event. Our proposal is based on a natural data-generating mechanism respecting that a disease score does not exist beyond the terminal event. We use modern semi-parametric statistical methods to provide robust and efficient estimation of the risk of terminal event and expected disease score conditional on no terminal event at a pre-specified landmark time. 
We also use the simultaneous asymptotic behavior of our estimators to develop a powerful closed testing procedure for confirmatory assessment of treatment effect on both onset of terminal event and level of disease score in the absence of a terminal event. A simulation study mimicking a large-scale outcome trial in chronic kidney patients as well as an analysis of that trial is provided to assess performance.
\end{abstract}

\noindent%
\vfill

\newpage
% \spacingset{1.8} % DON'T change the spacing!

%%%%%%%%%%%%%%%%%%%%%%%%%%%%%%%%%%%%%%%%%%%%%%%%%%

\section{Introduction}\label{sec:intro}

Clinical scores of organ conditions or physical ability - such as the estimated Glomerular Filtration Rate (eGFR) for
kidney function, Kansas City Cardiomyopathy Questionnaire (KCCQ) for heart
failure, or The Montreal Cognitive Assessment (MoCA) for cognitive decline - are
primary tools for evaluating treatment efficacy, yet they lose all clinical
meaning following terminal events like death or organ failure.
% Clinical scores of organ conditions or physical ability are not meaningful
% beyond events such as organ replacement therapy or death.
Consequently, in
trials where such terminal events are prevalent, this should be reflected by
statistical methods that target the impact of the treatment intervention on
disease scores. From a statistical point of view this poses a fundamental
challenge since effect of treatment cannot be sensibly quantified for the
clinical score in isolation.

The challenge of defining and quantifying treatment effects on outcomes that are
only meaningful in the absence of a terminal event, a phenomenon often referred
to as truncation by death, has received considerable attention and led to the
development of specialized causal inference approaches.
One prominent class of estimands is principal stratum effects, such as the
Survivor Average Causal Effect (SACE)
\citep{robins1986:sace,frangakis2002:prinicpalstratification},
% \citep{TchetgenTchetgen2014}
which attempts to quantify treatment effects on an outcome among the
unobservable subpopulation of individuals who would have avoided the terminal
event (e.g., survived) regardless of the treatment received. While SACE offers a
theoretically sound causal comparison by defining effects within a fixed, common
subpopulation, it remains questionable how SACE can inform treatment decisions or policies as the subpopulation cannot be identified or characterized in practice, may not
even exist, and its identification often relies on unfalsifiable assumptions
\citep{sjolander2011_principalstrat}. The practical utility in a clinical context,
as the one we consider here, is therefore limited.

Another approach is to consider treatment intervention in a scenario where the
terminal event can been prevented in the whole target population. If, in
reality, this is not feasible another strategy needs to be considered
\citep{Kahan2020}. In most settings concerning terminal events this approach is
indeed hypothetical.

More recently, separable effects have been proposed as an alternative
\citep{stensrud2023:conditionalseparableeffects, stensrud2022:ice_estimands},
aiming to quantify causal effects of modified versions of the study treatment
within an observable subset of the population. These effects, including
separable direct and indirect effects, attempt to disentangle causal mechanisms
by considering hypothetically modified treatments that operate through distinct
causal pathways. While addressing some limitations of SACE, this approach still
depends on the conceptualization of imagined interventions (i.e., modified
treatments and their isolation conditions) which may be hard to justify in
practice. From a clinical transparency perspective, relying on such hypothetical interventions to define a treatment effect can therefore be challenging for practical decision-making, especially in the primary evaluation of treatment efficacy. 

% Rather such an assessment needs to be balanced by the occurrence of terminal events either by:
% \begin{itemize}
% \item A joint assessment of occurrence of terminal event and score in the absence of terminal event
% \item An imagined intervention on terminal events enabling evaluation of scores in a hypothetical scenario where absence of terminal events is ensured
% \item An assessment on a subpopulation where the absence of terminal events are ensured.
% \end{itemize}

The problem of quantifying treatment effects using a clinical score that only exists in the absence of a terminal event, presents a substantial challenge in clinical trials. In such scenarios, the clinical score cannot be disentangled from the occurrence of the
terminal event and this needs to be reflected in the data-generating mechanism.

%% New part starts here
For full clinical transparency, the treatment effect should be expressed by
looking at what would have happened to this joint outcome under different
treatment options. One common strategy to simplify this problem is to combine
the terminal event and the clinical score into a single "utility'' measure by
assigning an arbitrary, unfavorable value to any patient who experiences a
terminal event. The choice of an unfavorable value is clearly a discussion point as it may ultimately govern conclusions about treatment effect \citep{Kurland2005} and also this summary is not fully informative about the probability of a terminal event.

To avoid the need for such arbitrary values, we propose a more informative
two-part summary measure in line with \citep{sjolander2011_principalstrat},
which evaluates the treatment’s benefit by simultaneously addressing two fundamental questions:
\begin{itemize}
  \item Under a given treatment intervention, what are the patient's chances of
  avoiding a terminal event?
  \item Under a given treatment intervention, what is the clinical condition (score) among patients who avoid the terminal event?
\end{itemize}
This dual assessment ensures that any reported improvement in a clinical score
is always balanced by the patient's likelihood of remaining event-free,
providing a transparent and comprehensive view of the treatment's impact.
Conceptually, this summary measure has previously been advocated in the context of the tripartite estimand framework due to its clinical relevance to patients, regulators and payers \citep{Akacha2017}.

It is important to recognize that the second clinical question—concerning the
clinical score only applies to a selected subgroup of patients: those who have
avoided a terminal event. One could naturally argue that comparing these scores
directly between treatment groups might lead to an unfair comparison, as the
treatment itself may influence the "selection" by determining which patients
survive or remain event-free.
As stated above and mirroring the developments
in \cite{Akacha2017}, our position is that this
conditional result should never be interpreted in isolation. Instead, the
clinical condition among survivors must always be offset by the likelihood of a
patient avoiding the terminal event in the first place. By utilizing the joint
summary measure proposed here, we facilitate a balanced and transparent
evaluation that accounts for these selection effects, ensuring the treatment's
impact on both survival and clinical status is considered simultaneously

Our proposal is focused around large scale randomized controlled trials in vulnerable populations where a surrogate marker along with a prevalent terminal event forms the basis of evaluating treatment effect. In particular we are motivated by the recently conducted FLOW trial \citep{Flow2024}. FLOW was a double-blind randomized controlled trial. The trial objective was to investigate the  ability of semaglutide - a once weekly glucagon like peptide-1 receptor agonist - to delay progression of kidney disease in a population with type 2 diabetes and chronic kidney disease at high risk of kidney disease progression.

A major challenge in this study was a substantial number of terminal events at any relevant landmark time after randomization \citep{Flow2024}. We will assess performance of our proposal in a simulation study mimicking the FLOW trial as well as analyse the actual trial data according to our proposal.

The paper is structured as follows. We introduce the formal set-up and define
the mathematical notation and the target parameters building on the above
exposition In Section \ref{sec:methods}. Assumptions and identifiability is addressed in
Section~\ref{sec:identification}, and
Section \ref{sec:est} is dedicated to describing randomization augmented
influence functions for these target parameters together with
estimators based on working prediction models for the nuisance components.
In Section~\ref{sec:extension} we describe extensions to general
covariate-dependent missing data and censoring problems.
A closed testing procedure for simultaneous assessment of effect on both the
disease score scale and the risk of terminal event is outlined in Section \ref{sec:testing}.
We present a Monte Carlo simulation study cast over the FLOW trial in Section \ref{sec:sim}
and proceed with an analysis of the FLOW data in Section \ref{sec:application}. Finally, a
discussion and directions for future research are outlined in Section \ref{sec:discussion}.

\section{Statistical Methods}\label{sec:methods}
\subsection{Setup and notation}\label{sec:setup}

We consider a setup where the occurrence of a terminal event that invalidates the
measurement of interest is recorded at some landmark time $\tau$ after randomization to treatment $A$. When such an event has not occurred prior to the landmark time the measurement of interest is meaningful and can be obtained at this landmark time.  

In this context, we denote the first occurrence of an event that invalidates the
measurement of interest by $T^\ast$. Furthermore, in the scenario $T^\ast\ge\tau$
where a meaningful clinical measurement of interest exists, we denote this
measurement by $Y$. For a treatment intervention $A=a$ we define the
counterfactual event-time $\potential{T^{\ast}}{a}$
% and $\potential{\epsilon^{\ast}}{a}$
as well as the counterfactual outcome $\potential{Y}{a}$ when $\potential{T^{\ast}}{a}\geq \tau$.

In terms of causal structures, the disease score and the terminal event are
inextricably linked because the score only exists and remains clinically
meaningful only when no event has occurred. Consequently, these cannot be
represented as separate, independent outcomes; instead, the treatment
intervention should be viewed as impacting a joint outcome consisting of both
the event status and the clinical score. When considering
a directed acyclic graphs the consequence is that the disease score, $Y$, in the absence of a terminal event, and  the occurrence of a terminal event, $T^\ast\geq\tau$, cannot naturally be represented as separate nodes. Instead, the
natural DAG relating treatment intervention, $A$, to $I(T^\ast\geq\tau)$ and $Y$, is simply
given by
\begin{align*}
\begin{tikzpicture}[>=latex, text height=1.5ex, text depth=0.25ex, ampersand replacement=\&]
  \matrix[row sep=1cm,column sep=0.5cm]{
    \node(a2) [] {$A$}; \& \& \node(z) []
    {$
    \begin{pmatrix}
    T^\ast\geq\tau\\
    Y
    \end{pmatrix}
    $}; \\
  };
  \path[->] (a2) edge[thick] node [auto] {} (z) ;
\end{tikzpicture}
\end{align*}

Based on the above DAG, we argue that for full clinical transparency the
treatment effect should be expressed in terms of joint counterfactuals (see
also \citep{sjolander2011_principalstrat})
$\{T^{\ast(a)}, Y^{(a)}\}$
that would have been realized under treatment intervention $A=a$, acknowledging
that this variable is not defined on the full product-space of the individual
components.
Taking this structural constraint seriously we propose a bivariate estimand that targets the two identifiable parts of the joint counterfactual distribution: the risk of the terminal event and the expected score conditional on remaining event-free across treatment interventions:
\begin{align}
  % &\estt{\theta}^{(a)} = \pr\left(\potential{T^{\ast}}{a}\leq\tau,\:\potential{\epsilon^{\ast}}{a}=1\right )\label{eq:targett},\\
  &\estt{\theta}^{(a)} = \pr\left(\potential{T^{\ast}}{a} \geq  \tau \right)\label{eq:targett},\\
  &\esty{\theta}^{(a)} = \E\left[\potential{Y}{a} \mid \potential{T^{\ast}}{a}\geq \tau\right] \label{eq:targety}.
\end{align}
The contrasts we consider in this context are given by:
\begin{align*}
&\estt{\psi} = \estt{\theta}^{(1)}-\estt{\theta}^{(0)},\\
&\esty{\psi} = \esty{\theta}^{(1)}-\esty{\theta}^{(0)}.
\end{align*}
Note that a positive value of $\estt{\psi}$ entails a reduction in the risk of
events that would prevent the measurement of interest at time $\tau$ due to
treatment. In addition, a positive value of $\esty{\psi}$ entails an increase in
the expected value of the clinical measurement at time $\tau$ among treated
patients with meaningful clinical measurement when comparing to comparator
treatment.

$\esty{\psi}$ should not be interpreted as a stand-alone and needs to be balanced by the chance of having a meaningful clinical measurement at time $\tau$, that is, by relating it to $P(\potential{T^{\ast}}{a}\geq \tau)$. We effectively achieve this by simultaneously considering $\estt{\psi}$ and $\esty{\psi}$ to gauge treatment effect.

Considering for instance chronic kidney disease, a drug is deemed beneficial if we can claim no
clinically relevant elevated risk of kidney failure or death due to treatment
% (under the implicit assumption that treatment does increase the risk of death from other causes)
and, in addition, an improvement in kidney function among the treated who are
still alive and have not had kidney failure at time $\tau$. If we formalize this statement it exactly corresponds to
simultaneously testing the two null-hypotheses
$$
\esty{H}: \esty{\psi}\leq \esty{\delta} \textnormal{ and }
\estt{H}: \estt{\psi}\leq -\estt{\delta}
$$
For some superiority margin $\esty{\delta}\geq0$ and some non-inferiority margin
$\estt{\delta}\geq0$. Note that for $\esty{\delta}=\estt{\delta}=0$ this corresponds to classical
testing for superiority of treatment. We revisit the testing procedures for this
testing problem in Section~\ref{sec:testing}.
\\[1em]

% Several distinct types of events may occur at this time and we denote the specific types
% by $\epsilon^{\ast}$. We assume that among these types of events, the event of interest
% for the evaluation of the treatment effect at $\tau$ is encoded as $\epsilon^{\ast}=1$.

In reality, the subjects in the trial may also drop-out at some time-point after
randomization either due to trial closeout or for other reasons. We denote this
censoring time by $C$ and let $T=T^{\ast}\wedge C \wedge \tau$ denote the first time either
censoring, a terminal event, or the measurement event occurs.
%where $\wedge$ denotes the minimum operator. 
We also let $\Delta=I(T^\ast\leq C)$ denote the indicator of whether censoring or an event is
observed.
% and let \(\epsilon = \epsilon^{\ast}\Delta\) be the observed cause of failure subject to right-censoring.
We note that in this scenario $Y$ may not be observed either due to censoring
before $\tau$, $(C<\tau)$, or if measurement is not obtained for other reasons. We let
$M$ be the indicator of whether $Y$ is observed ($M=1$) or not ($M=0$).
Similarly to clinical score this missing data mechanism is truncated by the
terminal event, and we define $R := I(T \geq \tau)M$.

% Given this interpretation, the benefits of a treatment intervention are
% naturally assessed by means of $\estt{\psi}$ and $\esty{\psi}$.
% This interpretation clearly comes with the cautionary note that due to competing
% risks, a reduction in risk of event of interest could potentially be caused by
% an increase in risk of competing events \cite{putter_2007}.

\subsection{Assumptions and Identification}\label{sec:identification}
In order to enable the assessment above we need to be able to
identify and estimate the targeted treatment contrasts from the observed data.
For this purpose, we further introduce a set of baseline covariates denoted by
$X$. To summarize, we observe
\begin{align*}
\{X, A, T, \Delta(\tau), \IT{} M,  \IT{} M Y\},
\end{align*}
where $T = \min(T^\ast, C, \tau)$ and
$\Delta(\tau) = I\{C \geq \min(T^{\ast}, \tau)\}$ for some censoring time $C$
and non-missingness $M$, and let $R := \IT{} M$.

% \begin{align*}
% \{X, A, T, \Delta(\tau), R,  R Y\},
% \end{align*}
% where $T = \min(T^\ast, C, \tau)$ and
% $\Delta(\tau) = I\{C \geq \min(T^{\ast}, \tau)\}$ for some censoring time $C$
% and non-missingness indicator $R$.

We proceed to formulate a set of missing data
assumptions that will enable identification in combination with standard
exchangeability and consistency assumptions. In addition, to allow for reliable
estimation, we are going to make a number of positivity assumptions and assume
that the randomized treatment is independent of the baseline covariates. Below, we list
the assumptions.
\begin{enumerate}[label=\emph{(A\arabic*)}]
  \item \label{as:randomization} Treatment randomization
\begin{align*}
    A \indep X
\end{align*}
\item  \label{as:exch} Exchangeability
\begin{align*}
  % \potential{Y}{a},\potential{T^{\ast}}{a}, \potential{\epsilon^{\ast}}{a} \indep A
  \potential{Y}{a},\potential{T^{\ast}}{a}, \potential{C}{a}, \potential{M}{a} \indep A
\end{align*}
\item \label{as:consistency} Consistency
\begin{align*}
    \potential{H^{}}{a} = H, H = (C,M,T^\ast, Y) \textnormal{ when } A=a
\end{align*}
\item \label{as:ypositivity} Positivity
\begin{align*}
  &\pr(A=a) > 0 \:\forall  a\\
  &\pr(T^\ast \geq \tau |A=a)>0 \:\forall  a
\end{align*}
\item \label{as:MCAR} Missing at random (outcome)
\begin{align*}
    Y\indep M \mid T^{\ast}\geq\tau, \:A
\end{align*}
\item \label{as:outcomecens} Random censoring (outcome)
\begin{align*}
  M,Y \indep I\{C\geq\tau\} \mid T^{\ast} \geq \tau, A
\end{align*}
\item \label{as:RandCens} Random censoring (time to event)
\begin{align*}
  % T^{\ast}, \epsilon^{\ast} \indep C   \mid A
  T^{\ast} \indep C   \mid A
\end{align*}
\item \label{as:cpositivity} Positivity (censoring/missingness)
\begin{align*}
  &\pr\left(C>\tau|A=a\right)>0 \:\forall  a\\
  &\pr\left(M=1|A=a, T^\ast \geq \tau \right)>0 \:\forall  a
\end{align*}
\end{enumerate}
Based on the above assumptions we are able to identify $\esty{\theta}^{(a)}$ from the
observed data through the expectation $\E\left[ I(A=a)\cdot R\cdot Y\right]$ and
$\pr(A=a, R=1)$ as follows:
\begin{align*}
\esty{\theta}^{(a)}&=\E\Big[\potential{Y}{a} \mid \potential{T^{\ast}}{a}\geq\tau
\Big]
\overset{\ref{as:exch}}{=}
\E\Big[\potential{Y}{a} \mid \potential{T^{\ast}}{a}\geq\tau, A=a\Big] \\
&\overset{\ref{as:consistency}}{=}
\E\Big[Y \mid
T^\ast \geq\tau, A=a\Big]
\overset{\ref{as:MCAR}, \ref{as:cpositivity}}{=}
  \frac{\E\Big[M Y \mid T^{\ast}\geq\tau,\:A=a\Big]}{\pr\Big(M=1 \mid T^{\ast}\geq\tau,\: A=a\Big)} \\
              &\overset{\ref{as:outcomecens}}{=}
                \frac{\E\Big[M Y \mid T\geq\tau,\:A=a\Big]}{\pr\Big(M=1 \mid T\geq\tau,\: A=a\Big)}
\overset{}{=}
\frac{\E\Big[I(A=a)\cdot R\cdot Y \Big]}{\pr\Big(A=a,R=1\Big)},
\end{align*}
where the last equality follows because $R=1$ entails $I(T\geq\tau)$.
Similarly, we are able to identify $\estt{\theta}^{(a)}$ from the observed data
as follows:
\begin{align*}
  \estt{\theta}^{(a)}&= \E\left[I(\potential{T^{\ast}}{a} \geq \tau) \right]\\
                     &\overset{\ref{as:exch}, \ref{as:consistency}}{=} \E\left[I(T^{\ast} \geq \tau) \mid A = a  \right] \\
                     &\overset{\ref{as:RandCens},\ref{as:cpositivity}}{=} \E\left[\frac{I(C\geq \tau)}{\pr(C \geq \tau|A=a)} I(T^{\ast} \geq \tau) \mid A = a  \right]\\
  &= \E\left[\frac{I(A=a)}{\pr(A=a)}\frac{\Delta(\tau)}{\pr(C \geq \tau|A=a)} I(T \geq \tau)  \right]
\end{align*}
As for the treatment randomization assumption, this is utilized in the next section, where we develop estimation procedures.

\subsection{Estimation procedure and asymptotics}\label{sec:est}
The estimation procedure developed in this section utilizes semi-parametric
theory to provide a model-agnostic framework for joint assessment. Unlike
traditional joint models that require explicit distributional or functional
assumptions (e.g., random slopes), our approach targets a bivariate estimand non-parametrically
through the projection of the empirical estimator influence function onto the tangent space
associated with the randomization assumption. While we employ specific working
models' — such as the Cox proportional hazards model for terminal events or
a linear model for the clinical score — these are used solely as nuisance
components. The resulting one-step estimator is consistent and asymptotically
normal even under misspecification of these working models, provided the
treatment randomization and missing data assumptions holds.
\\[1em]

Under the missing at random assumption \ref{as:MCAR}, a consistent estimator
of
$\esty{\theta}^{(a)} = \E[\potential{Y}{a} \mid \potential{T^{\ast}}{a}\geq \tau]$
can be obtained as 
\begin{align*}
  \esty{\widetilde{\theta}}^{(a)} = \frac{\sum_{i=1}^{n}I(A_{i}=a, R_{i}=1)Y}
  {\sum_{i=1}^{n} I(A_{i}=a, R_{i}=1)}.
\end{align*}
where $(A_{i}, R_{i}, R_{i}Y_{i}), i=1,\ldots,n$ are i.i.d. observations. The influence
function associated with this empirical estimator is given by
$\esty{\widetilde \phi}^{(a)}(Z; P_0)$, where
\begin{align*}
\esty{\widetilde \phi}^{(a)}(Z; P) = \frac{I(A=a)I(R=1)}{\pi(a)\rho(a)}\left\{Y - \esty{\theta}^{(a)}(P)\right\},
\end{align*}
and $\pi(a) = \pr_P(A=a)$, $\rho(a) = \pr_P(R=1|A=a)$.

With additional information on baseline covariates, this initial estimator can
be further
improved by exploiting the independence structure between the baseline covariates and the treatment due
to randomization \ref{as:randomization}.
The randomization augmented influence function (see Supplementary Material Section \ref{sec:eif:thetay})
is given by
\begin{align}\label{eq:eiffull}
  \begin{split}
  \esty{\phi}^{(a)}(Z; P)
  &= \frac{I(A=a)I(R=1)}{\pi(a)\rho(a)}\left\{Y - \esty{\theta}^{(a)}(P)\right\} \\
  &\qquad + \frac{\pi(a)-I(A=a)}{\pi(a)}\frac{\rho_{a}(X,a)}{\rho(a)} \left\{Q(X,a)-\esty{\theta}^{(a)}(P)\right\}
    ,
  \end{split}
\end{align}
with  \(Q(X,a) = \E_{P}[Y\mid X, A=a, R=1]\), and
\(\rho(X,a) =  \pr_P(R=1 \mid X, A=a)\).
To improve the efficiency of the initial empirical estimator we proceed by constructing a \emph{one-step estimator}
\citep{hines2022} in the following way. Let
\(\widehat{\mathcal{Q}} := \{\widehat{Q}(\cdot, a), \widehat{\rho}(\cdot, a), \widehat{\rho}(a), \widehat{\pi}(a), \esty{\widetilde{\theta}}^{(a)} \mid a=0,1\}\)
be estimates obtained from the observed data
where the estimators for the last three terms can be estimated consistently non-parametrically, and
the two first components are obtained as
predictions from some regression models. The initial estimate of
\(\esty{\theta}^{(a)}\)
can now be improved by adding the debiasing term derived from the plugin
estimate of the randomization augmented influence function
\begin{align*}
  \esty{\widehat{\theta}}^{(a)} = \esty{\widetilde{\theta}}^{(a)} +
  \Pn \esty{\phi}^{(a)}(Z; \widehat{\mathcal{Q}}),
\end{align*}
where we use the notation $\Pn$ to denote the empirical mean over the
i.i.d. observed data $Z_{1},\ldots,Z_{n}$ but keeping $\widehat{\mathcal{Q}}$ fixed. The
randomization of the treatment \(A\) guarantees that this estimator is
consistent irrespective of how we model the conditional means \(Q(X,a)\) and
\(\rho(X,a)\). Let \(Q^\ast(X,a)\) and
\(\rho^\ast(X,a)\) denote the probability limits for the chosen regression
models. Then under mild regularity conditions (see
Supplementary Material Section \ref{sec:app:asymp}) it holds that
\begin{align*}
  \sqrt{n}\{\esty{\widehat{\theta}}^{(a)} - \esty{\theta}^{(a)}\} &=
\frac{1}{\sqrt{n}}\sum_{i=1}^{n} \esty{\xi}^{(a)}(Z_{i}; \mathcal{Q}^{\ast}) + o_P(1)                                                                    
\end{align*}
where
\begin{align*}
  \esty{\xi}^{(a)}(Z; \mathcal{Q}^\ast)
  =
  \esty{\phi}^{(a)}(Z_{i}; \mathcal{Q}^\ast) +
\frac{I(A=a) - \pi_0(a)}{\pi_0(a)} \E\left[\frac{\rho^\ast(X,a)}{\rho_0(a)}\left\{Q^\ast(X,a)-\esty{\theta}^{(a)}\right\}\right].
\end{align*}

The joint distribution of
\((\esty{\widehat{\theta}}^{(1)}, \esty{\widehat{\theta}}^{(0)})^{\top}\)
follows directly from stacking the two influence functions
\begin{align*}
  \sqrt{n}\left\{
  \begin{pmatrix}
    \esty{\widehat{\theta}}^{(1)} \\
    \esty{\widehat{\theta}}^{(0)}
  \end{pmatrix} -
  \begin{pmatrix}
    \esty{{\theta}}^{(1)} \\
    \esty{{\theta}}^{(0)}
  \end{pmatrix}\right\} =
  \frac{1}{\sqrt{n}}\sum_{i=1}^{n}
  \begin{pmatrix}
    \esty{\xi}^{(1)}(Z_{i}; \mathcal{Q}^{\ast}) \\
    \esty{\xi}^{(0)}(Z_{i}; \mathcal{Q}^{\ast})
  \end{pmatrix} + o_{P}(1),
\end{align*}
which converges weakly to a Gaussian
with asymptotic variance that can be approximated by
\begin{align*}
  \widehat{\Sigma} = \frac{1}{n}\sum_{i=1}^{n}
  \begin{pmatrix}
    \esty{\xi}^{(1)}(Z_{i}; \widehat{\mathcal{Q}})^{2} &
\esty{\xi}^{(0)}(Z_{i}; \widehat{\mathcal{Q}})
\esty{\xi}^{(1)}(Z_{i}; \widehat{\mathcal{Q}}) \\
    \esty{\xi}^{(0)}(Z_{i}; \widehat{\mathcal{Q}})
\esty{\xi}^{(1)}(Z_{i}; \widehat{\mathcal{Q}}) &
\esty{\xi}^{(0)}(Z_{i}; \widehat{\mathcal{Q}})^{2} \\
    \end{pmatrix}.
\end{align*}
Finally, the estimate for
$\esty{\psi} = \esty{{\theta}}^{(1)} - \esty{{\theta}}^{(0)}$, is obtained as
\begin{align*}
  \esty{\widehat{\psi}} = \esty{\widehat{\theta}}^{(1)} - \esty{\widehat{\theta}}^{(0)}
\end{align*}
with the asymptotic variance approximated by
$(1 -1)\widehat{\Sigma}(1 -1)^{\top}$ and estimated influence function given by
\begin{align*}
  \esty{\xi}^{(1)}(Z_{i}; \widehat{\mathcal{Q}}) -
  \esty{\xi}^{(0)}(Z_{i}; \widehat{\mathcal{Q}}).
\end{align*}
Similarly, an efficient estimate of
 $\estt{\theta}^{(a)}$ can be obtained from an augmented influence function,
 % (Supplementary Material Section \ref{sec:eif:thetat})
 and  combined in a similar fashion into an estimate, $\estt{\widehat{\psi}}$ of the target parameter
 $\estt{\psi} = \estt {\theta}^{(1)} - \estt{\theta}^{(0)}$.
The details of this estimation procedure are given in more details in \citep{blanche22:binreg}
and are implemented in the R function \texttt{mets::binregATE}
\citep{metsr}. With access to the influence functions for both  $\estt{\widehat{\psi}}$ and
$\esty{\widehat{\psi}}$ we can use the stacking method above to calculate the
joint asymptotic distribution and correlation between the estimates that
we need for applying the multiple testing procedure that we describe in details
in the next section.
% For the utility $U^{(a)}$ we note that $\E[U^{(a)}] = \esty{\theta^{(a)}} S^{(a)}(\tau) + \Gamma\{1-S^{(a)}\}$
% where $S^{(a)} = \pr(T^{\ast}>\tau\mid A=a)$ of which we can constructor an
% estimator similarly to $\estt[]{\theta}^{(a)}$, and the influence function and
% asymptotic variance of the estimator for $\E[U^{(1)}]-\E[U^{(0)}]$ is then
% derived using the Delta method as above.
The estimators are implemented in the \texttt{targeted} R package \citep{targetedr} and
implementation and theoretical details are given in the Supplementary Material Section \ref{sec:software}.

In Section~\ref{sec:testing}, we transition from estimation to formal inference.
To provide a clinically transparent assessment, we must simultaneously evaluate
the impact of treatment on both the terminal event and the clinical score. We
utilize the joint asymptotic Gaussian distribution of the estimators derived in
this section to perform a closed testing procedure. This procedure ensures
Family-Wise Error Control at the $\alpha$ level: we first evaluate the
intersection hypothesis sing a signed Wald test, which incorporates the
estimated correlation between our targets to maximize power. Only if this
intersection is rejected do we proceed to test the individual hypotheses. The
use of a signed version of the Wald test is essential here to accommodate the
one-sided nature of clinical superiority and non-inferiority claims.

\subsubsection{Generalization to Covariate-Dependent Missingness}\label{sec:extension}

The methodology developed in the preceding sections provides a robust framework
for joint assessment without requiring explicit assumptions regarding
distribution of the clinical score or the specific timing of the
terminal event. However, the identification results in Section~\ref{sec:identification}
and the
estimators in Section~\ref{sec:est} hinge on assumptions
\ref{as:MCAR}, \ref{as:outcomecens}, and \ref{as:RandCens}
which treat missing clinical scores and censoring as independent of the outcomes.

While diagnostic models fitted to the missing data mechanisms in the FLOW study
studied in Section~\ref{sec:application} provided no indication that these
assumptions were not adequate, there is a clear
theoretical need to generalize the procedure for broader application. In the
following we therefore develop a natural extension of the framework that
accommodate a more complex set of missing data assumptions.
% In order to enable the assessment above we need to be able to
% identify and estimate the targeted treatment contrasts from the observed data.
% For this purpose, we further introduce a set of baseline covariates denoted by
% $X$. To summarize, we observe
% \begin{align*}
% \{X, A, T, \Delta(\tau), \IT{} R,  \IT{} R Y\},
% \end{align*}
% where $T = \min(T^\ast, C, \tau)$ and
% $\Delta(\tau) = I\{C \geq \min(T^{\ast}, \tau)\}$ for some censoring time $C$
% and non-missingness $R$.
% We proceed to formulate a set of missing data
% assumptions that will enable identification in combination with standard
% exchangeability and consistency assumptions. In addition, to allow for reliable
% estimation, we are going to make a number of positivity assumptions and assume
% that the randomized treatment is independent of the baseline covariates. Below, we list
% the assumptions.

In the general setting, we replace assumptions \ref{as:MCAR},
\ref{as:outcomecens}, \ref{as:RandCens}, and \ref{as:cpositivity} with
\begin{enumerate}[label=\emph{(A\arabic*')}]
\setcounter{enumi}{4}
\item \label{as:MAR_ex} Missing at random (outcome)
\begin{align*}
  Y \indep M \mid T^{\ast} \geq \tau,A, X
\end{align*}
\item \label{as:outcomecens_ex} Random censoring (outcome)
\begin{align*}
  M,Y \indep I\{C\geq\tau\} \mid T^{\ast} \geq \tau, A, X
\end{align*}
\item \label{as:RandCens_ex} Random censoring (time to event)
\begin{align*}
  % T^{\ast}, \epsilon^{\ast} \indep C   \mid A
   T^{\ast} \indep C   \mid A, X
\end{align*}
\item \label{as:cpositivity_ex} Positivity (censoring/missingness)
\begin{align*}
  &\pr\left(M=1|T \geq \tau, A=a,X=x\right)>0 \:\forall  a, x\\
  &\pr\left(C>\tau|A=a, X = x\right)>0 \:\forall  a, x
\end{align*}
\end{enumerate}
Based on the above assumptions we are able to identify $\esty{\theta}^{(a)}$ from the
observed data as follows:
\begin{align*}
  &\E[\frac{I(A = a) \IT{} M Y}{\pr(A=1)\pr(T\geq\tau|A=a, X)\pr(M = 1 | T\geq\tau, A = a, X)}]\\
  \overset{\ref{as:consistency}}{=}& \E\left [  \frac{I(A = a) \IT{(a)} \potential{M}{a} \potential{Y}{a}}{\pr(A=1)\pr(T\geq\tau|A=a, X)\pr(M = 1 | T\geq \tau, A = a, X)}  \right ] \\
    \overset{\ref{as:randomization},\ref{as:exch}}{=}& \E\left [  \frac{I\{C^{(a)}\geq\tau\} \IT{*(a)} \potential{M}{a} \potential{Y}{a}}{\pr(T\geq\tau|A=a, X)\pr(M = 1 | T\geq\tau, A = a, X)}  \right]\\
 \overset{\ref{as:outcomecens_ex}, \ref{as:RandCens_ex}}{=}& \E\left[ \frac{\E\left[I\{C^{(a)}\geq\tau\}|X\right] \E\left[\IT{*(a)} \potential{M}{a} \potential{Y}{a}|X\right]}{\pr(\potential{C}{a}>\tau |X)\pr(\potential{T^{\ast}}{a}>\tau |X)\pr(\potential{M}{a} = 1 | \potential{T^\ast}{a}>\tau, X)}  \right] \\
  \overset{\ref{as:MAR_ex}}{=}& \E\left [  \frac{\E[\potential{M}{a}|\potential{T}{a} \geq \tau, X] \E[\IT{(a)}  \potential{Y}{a}|X]}{\pr(\potential{M}{a} = 1 | \potential{T^\ast}{a}\geq\tau, X)\pr(\potential{T^{\ast}}{a}\geq\tau |X)}  \right]\\
  =&  \E\left [ \frac{ \E[I(\potential{T^{\ast}}{a}\geq\tau)  \potential{Y}{a}|X]}{\pr(\potential{T^{\ast}}{a}\geq\tau |X)} \right]\\
  =& \E[\potential{Y}{a}|\potential{T^{\ast}}a \geq \tau].
\end{align*}
The identification of
$\estt[]{\theta}^{(a)}$ follows
from a similar argument as in Section~\ref{sec:identification}.

\subsubsection*{Estimation procedure}

Under minimal nuisance model convergence assumptions we want to construct a consistent RAL estimator of
$\esty{\theta}^{(a)} = \E[\potential{Y}{a} \mid \potential{T^{\ast}}{a}\geq \tau]$.
As shown in Supplementary Material Section~\ref{sec:gen-eif}, the efficient influence function (EIF)
for this target parameter is
\begin{align*}
  &\esty{\phi}^{(a)}(Z; P)   =\,\frac{I(A=a)}{\pi(a)}\frac{\Delta(\tau)}{S_C(\tau \mid X,A)} \frac{\IT{}}{S(a)} \frac{R}{\rho(X,A)} \left\{Y - \esty{\theta}^{(a)}(P)\right\} \\
  &\qquad +\,  \frac{I(A=a)}{\pi(a)}\frac{\Delta(\tau)}{S_C(\tau|X,A)} \frac{\IT{}}{S(a)} \frac{\rho(X,A)-R}{\rho(X,A)}\left\{Q(X,A) - \esty{\theta}^{(a)}(P)\right\}\\
  &\qquad+\, \frac{I(A=a)}{\pi(a)} \int_{0}^\tau \frac{1}{S(u|X, A)S_C(u-\mid X,A)}\,dM_{C}(u\mid X,A)\frac{S(X,a)}{S(a)}\left\{Q(X, a)-\esty{\theta}^{(a)}(P)\right\} \\
  &\qquad+\,  \frac{\{\pi(a) - I(A = a)\}}{\pi(a)}\frac{S(X,a)}{S(a)}\left\{Q(X, a)-\esty{\theta}^{(a)}(P)\right\},
\end{align*}
with
\begin{align*}
  \pi(a) &= \pr_{P}\left(A=a \right)\\
  S_C(u|X, A) &= \pr_{P}\left(C \geq u |X,A \right)\\
  S(u|X, A) &= \pr_{P}\left(T^\ast \geq u |X,A \right)\\
  S(X, A) &= \pr_{P}\left(T^\ast \geq \tau |X,A \right)\\
  S(A) &= \pr_{P}\left(T^\ast \geq \tau |A \right)\\
  \rho(X,A) &=  \pr_P(R=1 \mid X, A, T^\ast \geq \tau)\\
  Q(x, a) &= \E_{P}\left[Y \mid X, A=a, T^\ast \geq \tau \right]\\
  M_C(u|X,A) &= (1-\Delta(\tau))I(T\leq u) - \int_0^{u\wedge \tau} I(T \geq u) d \Lambda_C(u|X,A)
\end{align*}
We note that the EIF, $P\mapsto \esty{\phi}^{(a)}(Z; P)$,
evaluated in the true probability distribution $P_0$
depends only on the probability distribution through the above functions
and $\esty{\theta}^{(a)}(P)$, which we collectively denote \(\mathcal{Q}(P)\).

Again, we proceed by constructing a one-step estimator
\citep{hines2022}. Let
\(\widehat{\mathcal{Q}}\)
be estimates obtained from the observed data.
An initial plug-in estimate for $\esty{\theta}^{(a)}$ is given by
\begin{align*}
  \esty{\widetilde{\theta}}^{(a)}(\widehat{\mathcal{Q}}) =
  \Pn \frac{I(A=a)}{\widehat \pi(a)}\frac{\Delta(\tau)}{\widehat S_C(\tau \mid X,A)} \frac{\IT{}}{\widehat S(a)} \frac{R}{\widehat \rho(X,A)} Y.
\end{align*}
The plug-in estimate is augmented by adding a debiasing term associated with the EIF
\begin{align*}
  \esty{\widehat{\theta}}^{(a)} = \esty{\widetilde{\theta}}^{(a)} +
  \Pn \esty{\phi}^{(a)}(Z; \widehat{\mathcal{Q}}).
\end{align*}

The asymptotic variance of this estimator can be consistently estimated from the
empirical variance of its influence function, facilitating valid statistical
inference. This property holds under mild regularity conditions, provided that
the nuisance models, which account for baseline covariates X, converge at a
sufficiently fast rate. Crucially, a parametric convergence rate for these
covariate-dependent nuisance models is not required. Instead, the estimator
achieves its properties if the product of the errors from the outcome nuisance
models and the censoring/missingness models converges at a rate of
$o_{P}(n^{-1/2})$ convergence rate in $L^2_{P}$. These requirements, which are
governed by the second-order remainder term detailed in the Supplementary
Material, ensure that the estimator remains robust and semi-parametrically
efficient even when flexible, non-parametric models or machine learning
techniques are employed.

It also follows from the second-order remainder term that the one-step estimator
exhibits double robustness properties, in the sense that the estimator maintains
its consistency even under partial model misspecification, provided that any of
the following conditions are met:
\begin{itemize}
  \item Both the outcome model $Q$ and the censoring model $S_C$ are correctly specified
  \item Both the missing data model $\rho$ and the censoring model $S_C$ are correctly
  specified
  \item Both the outcome model $Q$ and the survival model $S$ are
correctly specified
\end{itemize}
This robustness ensures that the joint assessment remains valid in complex
settings where it may be difficult to perfectly specify every aspect of the
data-generating mechanism, such as the exact relationship between baseline
characteristics and the clinical score or the precise timing of terminal events.

% The one-step estimator will be asymptotically linear under further
% convergence rate assumptions of the nuisance models depending on the baseline
% covariates $X$. The required assumptions are evident from the second order
% remainder, which is stated in Supplementary Material \ref{sec:gen-eif}.
% Importantly, a parametric convergence rate of the covariate depended nuisance
% models are not required. Instead we require a $o_{P}(n^{-1/2})$ convergence rate
% in $L^2_{P}$ on product structures of the outcome nuisance models and the
% censoring and missingness models.

The derivation of the EIF for the terminal event risk $\estt[]{\theta}^{(a)}$, under the generalized assumptions of covariate-dependent censoring \ref{as:RandCens_ex}, and
the corresponding positivity requirement \ref{as:cpositivity_ex} has been
studied extensively in the field.
Detailed results and theoretical properties for these estimators can be
found in, for example,
\citep{hubbard2000:survtreat}, \citep{blanche22:binreg}, and
\citep{westling2024inference}. The corresponding semi-parametric efficient estimator is also double-robust in
the sense that it remains consistent if either the survival function $S$ of the
terminal event or the censoring distribution $S_C$ is correctly specified.
Implementations for these models are readily
available in the R statistical environment through the packages \texttt{mets}
\citep{metsr} and
\texttt{riskRegression}
\citep{ozenne2020:atecens}.

\subsection{A closed testing procedure based on signed Wald tests}\label{sec:testing}

In order to provide family-wise error control at $\alpha$ level when simultaneously evaluating $H_{Y|T^\ast}$ and $H_{T^\ast}$ we propose a closed testing procedure in which $H_{Y|T^\ast}\cap H_{T^\ast}$ is evaluated with an $\alpha$ level test and, contingent on the rejection of the intersection hypothesis, $H_{Y|T^\ast}$ and $H_{T^\ast}$ are evaluated separately, also by $\alpha$ level tests  \citep{closedtesting}.

To efficiently test the intersection hypothesis at $\alpha$ level we consider a Wald test proposed in for instance \citep[p. 224]{robertson1988order} or \citep{silvapulle92_robustwald} for general-purpose hypothesis testing. In our particular context, we consider a version of this test that is truncated at zero for values below zero, and we term this the signed Wald test in what follows. Accordingly, the signed Wald test for testing $H_{Y|T^\ast}\cap H_{T^\ast}$ is defined as follows:
$$
SW_{n,H_{Y|T^\ast}\cap H_{T^\ast}}=\inf_{\psi\in H_{Y|T^\ast}\cap H_{T^\ast}}\big\{n\cdot\{\hat{\psi}-\psi\}^{\top}\hat{\Sigma}^{-1}\{\hat{\psi}-\psi\}\big\},
$$
where $\psi=\{\psi_{Y | T^{\ast}},\psi_{T^{\ast}} \}^{T}$ and $\hat{\psi}=\{\hat{\psi}_{Y | T^{\ast}},\hat{\psi}_{T^{\ast}} \}^{T}$.

\begin{figure}[htpb]
  \centering
  \includegraphics[width=0.7\textwidth]{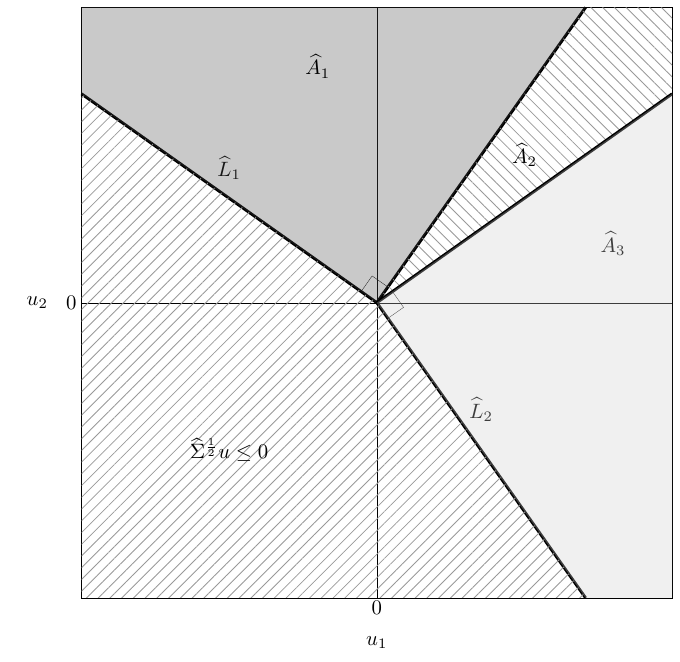}
  \caption{Regions characterizing the value of the signed Wald test}
  \label{fig:wald}
\end{figure}

In order to derive large sample properties of  $SW_{n,H_{Y|T^\ast}\cap H_{T^\ast}}$ we first rewrite above expression in terms of $\hat{u}=\sqrt{n}\cdot\sqrt{\hat{\Sigma}^{-1}}\{\hat{\psi}-(\esty{\delta},-\estt{\delta})^{\top}\}$ and $u=\sqrt{n}\cdot\sqrt{\hat{\Sigma}^{-1}}\{\psi-(\esty{\delta},-\estt{\delta})^{\top}\}$ to obtain:
\begin{equation}\label{SignWald}
SW_{n,H_{Y|T^\ast}\cap H_{T^\ast}}=\inf_{\sqrt{\hat{\Sigma}}u\leq 0} \big\{\{\hat{u}-u\}^{\top}\{\hat{u}-u\}\big\}=\inf_{\sqrt{\hat{\Sigma}}u\leq 0} \|\hat{u}-u\|^{2}
\end{equation}
As illustrated in Figure \ref{fig:wald} the region $\{u: \sqrt{\hat{\Sigma}}u\leq0\}$ is enclosed by the two lines $\hat{L}_{1}$ and $\hat{L}_{2}$. Note that if $\hat{u}$ belongs to that  region the signed wald test equals zero. If $\hat{u}\in\hat{A}_{1}$ we know that the projection of $\hat{u}$ onto $\hat{L}_{1}$ is the point in  $\{u: \sqrt{\hat{\Sigma}}u\leq0\}$ closest to $\hat{u}$. Accordingly, for $\hat{u}\in\hat{A}_{1}$, we have $SW_{n,H_{Y|T^\ast}\cap H_{T^\ast}}=\|\hat{u}-P_{\hat{L}_{1}}(\hat{u})\|^{2}$, where $P_{\hat{L}_{1}}(\hat{u})$ denotes the projection of $\hat{u}$ onto $\hat{L}_{1}$. Similarly it follows that $SW_{n,H_{Y|T^\ast}\cap H_{T^\ast}}=\|\hat{u}-P_{\hat{L}_{2}}(\hat{u})\|^{2}$ for $\hat{u}\in\hat{A}_{3}$. Finally, for $\hat{u}\in\hat{A}_{2}$ the point in  $\{u: \sqrt{\hat{\Sigma}}u\leq0\}$ closest to $\hat{u}$ is zero and accordingly $SW_{n,H_{Y|T^\ast}\cap H_{T^\ast}}=\|\hat{u}\|^{2}$ in this case.

In summary, we conclude that the signed Wald test for $H_{Y|T^\ast}\cap H_{T^\ast}$ may be rewritten as:  
$$
SW_{n,H_{Y|T^\ast}\cap H_{T^\ast}}=I(\hat{u}\in\hat{A}_{1})\cdot\|\hat{u}-P_{\hat{L}_{1}}(\hat{u})\|^{2}+ I(\hat{u}\in\hat{A}_{3})\cdot\|\hat{u}-P_{\hat{L}_{2}}(\hat{u})\|^{2}+
I(\hat{u}\in\hat{A}_{2})\cdot\|\hat{u}\|^{2}
$$

Next note that when $\psi=(\esty{\delta},-\estt{\delta})^{\top}$ we have that $\hat{u}$ converges weakly to a zero mean standard normal distribution. We also have that $\hat{\Sigma}$ converges in probability to some positive definite matrix $\Sigma$. It follows from the above representation of $SW_{n,H_{Y|T^\ast}\cap H_{T^\ast}}$ that for $\psi=(\esty{\delta},-\estt{\delta})^{\top}$:
\begin{equation}\label{mixdist}
SW_{n,H_{Y|T^\ast}\cap H_{T^\ast}}\rightsquigarrow \left(\frac{1}{2}-q \right)\cdot\chi^{2}_{0}+\frac{1}{2}\cdot\chi^{2}_{1}+q\cdot\chi^{2}_{2}
\end{equation}
where $q=P(\varepsilon\in A_{2})$, with
$\varepsilon\sim N(0,I_{2\times2})$ and $A_{2}$ defined as $\hat{A}_{2}$ when replacing $\hat{\Sigma}$ by $\Sigma$.
It follows that the p-value, that, is the maximal tail probability in the distribution of $SW_{n,H_{Y|T^\ast}\cap H_{T^\ast}}$ under the null hypothesis, can be approximated as 
\begin{eqnarray*}
  &\sup_{\psi\in H_{Y|T^\ast}\cap H_{T^\ast}}P_{\psi}(SW_{n,H_{Y|T^\ast}\cap H_{T^\ast}}\geq x)&=P_{\psi=(\esty{\delta},-\estt{\delta})^{\top}}(SW_{n,H_{Y|T^\ast}\cap H_{T^\ast}}\geq x) \\
  & &\quad \longrightarrow P(SW_{H_{Y|T^\ast}\cap H_{T^\ast}}\geq x), as\:\: n\rightarrow\infty
\end{eqnarray*}
where
$$
SW_{H_{Y|T^\ast}\cap H_{T^\ast}}\sim \left(\frac{1}{2}-q\right)\cdot\chi^{2}_{0}+\frac{1}{2}\cdot\chi^{2}_{1}+q\cdot\chi^{2}_{2}.
$$
To calculate the p-value in practice based on the above approximation we also
need to consistently estimate $q$ and plug the resulting estimator into the
right-hand side of (\ref{mixdist}). Such an estimator is obtained by noting that $P(\sqrt{\Sigma}\varepsilon\leq 0)=\frac{1}{2}-q$.
It follows that we can consistently estimate $q$ by 
$
\hat{q}=\frac{1}{2}-P(\sqrt{\hat{\Sigma}}\varepsilon\leq 0).
$
Here we note that $P(\sqrt{\hat{\Sigma}}\varepsilon\leq 0)$ is easy to calculate by either simulation or numerical integration.

%Also in order to more easily calculate $SW_{n,H_{Y|T^\ast}\cap H_{T^\ast}}$ in practice we re-express this quantity in terms of the standardized estimates %$z_{Y|T^{\ast}}=\frac{\sqrt{n}\cdot(\hat{\psi}_{Y | T^{\ast}}-\esty{\delta})}{\sqrt{\hat{\Sigma}_{11}}}$, %$z_{T^{\ast}}=\frac{\sqrt{n}\cdot(\hat{\psi}_{T^{\ast}}+\estt{\delta})}{\sqrt{\hat{\Sigma}_{22}}}$, and the estimated correlation %$\hat{\rho}=\frac{\hat{\Sigma}_{12}}{\sqrt{\hat{\Sigma}_{11}\cdot \hat{\Sigma}_{22} }}$ between the estimators . By brute force calculation we obtain:

%\begin{eqnarray*}
% &&SW_{n,H_{Y|T^\ast}\cap H_{T^\ast}}=I\big(z_{Y|T^{\ast}}\geq 0,\: z_{T^{\ast}}\leq\hat{\rho}\cdot z_{Y|T^{\ast}}\big)\cdot %z_{Y|T^{\ast}}^{2}+I\big(z_{T^{\ast}}\geq0,\: z_{Y|T^{\ast}}\leq \hat{\rho}\cdot z_{T^{\ast}}\big)\cdot z_{T^{\ast}}^{2}\nonumber\\
% &&+I\big(z_{Y|T^{\ast}}\geq 0,\: z_{T^{\ast}}\geq\hat{\rho}\cdot z_{Y|T^{\ast}} \cup z_{T^{\ast}}\geq0,\: z_{Y|T^{\ast}}\geq \hat{\rho}\cdot z_{T^{\ast}}  %\big)\frac{z_{Y|T^{\ast}}^{2}+z_{T^{\ast}}^{2}-2\cdot\hat{\rho}\cdot z_{Y|T^{\ast}}\cdot z_{T^{\ast}}}{1-\hat{\rho}^{2}}
%\end{eqnarray*}
%which is straightforward to calculate by plug in.

For testing the single hypotheses $H_{Y| T^{\ast}}$ and $H_{T^{\ast}}$ we again use signed Wald tests which are the standard testing tool for single parameter superiority/non-inforiority testing. Specifically, with $z_{Y|T^{\ast}}=\frac{\sqrt{n}\cdot(\hat{\psi}_{Y | T^{\ast}}-\esty{\delta})}{\sqrt{\hat{\Sigma}_{11}}}$, $z_{T^{\ast}}=\frac{\sqrt{n}\cdot(\hat{\psi}_{T^{\ast}}+\estt{\delta})}{\sqrt{\hat{\Sigma}_{22}}}$ denoting the standardized estimates, the single hypothesis signed Wald tests are given by:
\begin{eqnarray*}
    &&SW_{n,H_{Y|T^{\ast}}}=I(z_{Y|T^{\ast}}\geq 0)\cdot z_{Y|T^{\ast}}^{2},\\
    &&SW_{n,H_{T^{\ast}}}=I(z_{T^{\ast}}\geq0)\cdot z_{T^{\ast}}^{2}.
\end{eqnarray*}

The accompanying p-values are computed by approximations similar to that of the intersection hypothesis test, that is: 
\begin{align*}
  \sup_{\psi\in H_{Y|T^{\ast}}}P_{\psi}(SW_{n,H_{Y|T^{\ast}}}\geq x) &= P_{\psi_{Y|T^{\ast}}=\esty{\delta}}(SW_{n,H_{Y|T^{\ast}}}\geq x) \\
                                                      &\quad \longrightarrow P(SW_{H_{Y|T^{\ast}}}\geq x), \text{
                                                        as}\:\: n\rightarrow\infty,\\
  \sup_{\psi\in H_{T^{\ast}}}P_{\psi}(SW_{n,H_{T^{\ast}}}\geq x) &= P_{\psi_{T^{\ast}}=-\estt{\delta}}(SW_{n,H_{T^{\ast}}}\geq x) \\
  &\quad \longrightarrow P(SW_{H_{T^{\ast}}}\geq x), as\:\: n\rightarrow\infty, \\
\end{align*}
where
\begin{eqnarray*}
  &&SW_{H_{Y|T^{\ast}}}\sim \frac{1}{2}\cdot\chi^{2}_{0}+\frac{1}{2}\cdot\chi^{2}_{1},\\
  &&SW_{H_{T^{\ast}}}\sim \frac{1}{2}\cdot\chi^{2}_{0}+\frac{1}{2}\cdot\chi^{2}_{1}.
\end{eqnarray*}

In the Supplementary Material Section~\ref{sec:supplpower} we show that when there is a substantial positive correlation between the estimated target parameters the proposal for simultaneously evaluating $H_{Y|T^\ast}$ and $H_{T^\ast}$ has higher disjunctive (reject at least one hypothesis) power than the Bonferroni-Holm procedure under any alternative. Moreover, the proposal has higher conjunctive (reject both hypotheses) power than the Bonferroni-Holm procedure in all correlation scenarios and under all alternatives. We also argue that in practice the power gains can be substantial.

\section{Simulation study}\label{sec:sim}

In order to rigorously assess the performance of our proposed estimators and
closed-testing framework in a realistic context, we have designed a
comprehensive Monte Carlo simulation study. This simulation has been
calibrated to mirror the characteristics of the FLOW trial \citep{Flow2024} to which we also apply the methodology later. The following variables are considered in this simulation study
\begin{description}
  \item[$T$:] time of first event in years (first major irreversible kidney event or non-related death).
  \item[$\epsilon$:] event type at $T$; first major irreversible kidney event ($\epsilon=1$), non-related death
        ($\epsilon=2$), or right censoring ($\epsilon=0$).
  \item[$Y := Y(\tau)$:] clinical outcome measurement (eGFR) at landmark time $\tau$.
  \item[$R$:] missing indicator for $Y$ (1 if observed, 0 if either $T<\tau$ or
        if $Y$ was not measured for other reasons).
  \item[$A$:] binary treatment (1: active, 0: placebo).
  \item[$X_{1}$:] covariate, clinical outcome at baseline (eGFR).
  \item[$X_{2}$:] covariate, binary treatment usage indicator (1: SGLT2 treatment, 0: none).
\end{description}
Let the covariates be distributed according to
$A \sim \operatorname{Bernoulli}(\pi)$,
$X_{2} \sim \operatorname{Bernoulli}(p_{X_{2}})$, and
$X_{1}|X_{2}=x \sim \mathcal{N}(\mu_{x}, \sigma_{x}^{2}), x\in\{0,1\}$.
The clinical outcome is modelled as
\begin{align*}
  Y\mid A, X_{1},X_{2} \sim \mathcal{N}(
  \beta_{Y,0}^{(A)} + \beta_{Y,1}^{(A)}(X_{1}-\mu_{1}) + \beta^{(A)}_{Y,2}X_{2},
  \sigma^{(A)}_{Y}{}^{2}
  ),
\end{align*}
which is observed conditional on the patients not experiencing a terminal event and staying in study until the
landmark time $\tau$, with the status described by $R$ ($R=1$ corresponds to actually observed). The status variable $R$ is modelled as
\begin{align*}
  R \mid T^{\ast}\geq\tau, A,X_{1},X_{2} \sim \operatorname{Bernoulli}\left(\operatorname{expit}\{\beta_{R,0}^{(A)} +
  \beta_{R,1}^{(A)}(X_{1}-\mu_{1}) + \beta_{R,2}^{(A)}X_{2}
  \}\right)
\end{align*}
The cause-specific hazard for all events and censoring are modelled as  Cox proportional hazard models with the baseline hazard function
described by a Weibull hazard function parametrized in the following way
\begin{align*}
  \lambda_{\epsilon=k}(t\mid A,X_{1},X_{2}) = \gamma_{\epsilon=k}^{(A)}t^{\gamma^{(A)}_{\epsilon=k}{}^{-1}}
  \exp\left\{\beta_{\epsilon=k,0}^{(A)} + \beta_{\epsilon=k,1}^{(A)}(X_{1}-\mu_{1}) + \beta_{\epsilon=k,2}^{(A)}X_{2}\right\}, k=0,1,2.
\end{align*}

\subsection{Simulation results}\label{sec:simres}

The parameters of the simulation study are calibrated to the FLOW study and are
defined in Table \ref{tab:simpar}. For the clinical outcome model we observe
strong effects of both $X_{1}$, and $X_{2}$. For the cause-specific hazards for
both first major irreversible kidney event and non-related death  
more modest statistical evidence of associations are seen. The censoring
distribution is almost entirely driven by administrative censoring and as a natural consequence we
do not see any statistical evidence of effects of the two covariates. The same
applies for the missing data mechanism conditioned on $T\geq\tau$ indicating that
the assumptions \ref{as:MCAR}, \ref{as:RandCens} are reasonable in this
application and accordingly we enforce these assumptions in the simulation scenarios (Table \ref{tab:simpar}). We consider the fixed landmark time $\tau=2$

\begin{table}
  \caption{Parameters of the simulation study.\label{tab:simpar}}
  \centering
\begin{tabular}{rrrrrrrrr}
  \toprule
  % \rowcolor{Gray}
  & $\bm{\pi}$ &  &  &  &  &  &  & \\
  \addlinespace
  $A$ & 0.5 &  &  &  &  &  &  & \\
  \midrule
  % \addlinespace
  % \rowcolor{Gray}
  & $\bm{\mu_{1}}$ & $\bm{\sigma_{1}}$ & $\bm{\mu_{2}}$ & $\bm{\sigma_{2}}$ &  &
                              &  & \\
  \addlinespace
  $X_{1}$ & 46.24 & 14.99 & 51.15 & 15.33 &  &  &  & \\
  \midrule
  % \addlinespace
  % \midrule
  % \rowcolor{Gray}
  & $\bm{p_{X_2}}$ &  &  &  &  &  &  & \\
  \addlinespace
  $X_{2}$ & 0.156 &  &  &  &  &  &  & \\
  \midrule
  \addlinespace
  % \midrule
  % \rowcolor{Gray}
 & $\bm{\beta_{Y,0}^{(A=0)}}$ & $\bm{\beta_{Y,1}^{(A=0)}}$ & $\bm{\beta_{Y,2}^{(A=0)}}$ & $\bm{\sigma_{Y}^{(A=0)}}$ & $\bm{\beta_{Y,0}^{(A=1)}}$ & $\bm{\beta_{Y,1}^{(A=1)}}$ & $\bm{\beta_{Y,2}^{(A=1)}}$ & $\bm{\sigma_{Y}^{(A=1)}}$\\
  \addlinespace
  $Y$ & 40.141 & 0.895 & 1.993 & 11.85 & 43.121 & 0.863 & 2.620 & 12.16\\
  \midrule
  \addlinespace
  % \midrule
  % \rowcolor{Gray}
 & $\bm{\beta_{R,0}^{(A=0)}}$ & $\bm{\beta_{R,1}^{(A=0)}}$ & $\bm{\beta_{R,2}^{(A=0)}}$ &  & $\bm{\beta_{R,0}^{(A=1)}}$ & $\bm{\beta_{R,1}^{(A=1)}}$ & $\bm{\beta_{R,2}^{(A=1)}}$ & \\
  % $R\mid T\geq\tau$ & 2.243 & 0.0049 & 0.431 &  & 2.309 & 0.00074 & 0.391 & \\
\addlinespace
$R\mid T\geq\tau$ & 2.243 & 0 & 0 &  & 2.309 & 0 & 0 & \\
  \midrule
  \addlinespace
  % \midrule
  % \rowcolor{Gray}
 & $\bm{\beta_{\epsilon=0,0}^{(A=0)}}$ & $\bm{\beta_{\epsilon=0,1}^{(A=0)}}$ & $\bm{\beta_{\epsilon=0,2}^{(A=0)}}$ & $\bm{\gamma_{\epsilon=0}^{(A=0)}}$ & $\bm{\beta_{\epsilon=0,0}^{(A=1)}}$ & $\bm{\beta_{\epsilon=0,1}^{(A=1)}}$ & $\bm{\beta_{\epsilon=0,2}^{(A=1)}}$ & $\bm{\gamma_{\epsilon=0}^{(A=1)}}$\\
\addlinespace
  $\epsilon=0$ & -8.874 & 0 & 0 & 6.691 & -9.278 & 0 & 0 & 6.946\\
  \midrule
  \addlinespace
  % \midrule
  % \rowcolor{Gray}
 & $\bm{\beta_{\epsilon=1,0}^{(A=0)}}$ & $\bm{\beta_{\epsilon=1,1}^{(A=0)}}$ & $\bm{\beta_{\epsilon=1,2}^{(A=0)}}$ & $\bm{\gamma_{\epsilon=1}^{(A=0)}}$ & $\bm{\beta_{\epsilon=1,0}^{(A=1)}}$ & $\bm{\beta_{\epsilon=1,1}^{(A=1)}}$ & $\bm{\beta_{\epsilon=1,2}^{(A=1)}}$ & $\bm{\gamma_{\epsilon=1}^{(A=1)}}$\\
  \addlinespace
  $\epsilon=1$ & -3.558 & -0.0243 & -0.583 & 1.822 & -4.008 & -0.0289 & -0.126 &
                                                                                 1.901\\
  \midrule
  \addlinespace
  % \midrule
  % \rowcolor{Gray}
 & $\bm{\beta_{\epsilon=2,0}^{(A=0)}}$ & $\bm{\beta_{\epsilon=2,1}^{(A=0)}}$ & $\bm{\beta_{\epsilon=2,2}^{(A=0)}}$ & $\bm{\gamma_{\epsilon=2}^{(A=0)}}$ & $\bm{\beta_{\epsilon=2,0}^{(A=1)}}$ & $\bm{\beta_{\epsilon=2,1}^{(A=1)}}$ & $\bm{\beta_{\epsilon=2,2}^{(A=1)}}$ & $\bm{\gamma_{\epsilon=2}^{(A=1)}}$\\
  \addlinespace
  $\epsilon=2$ & -4.173 & -0.0205 & -0.455 & 1.143 & -4.135 & 0.00687 & -0.598 &
                                                                                 1.071\\
\bottomrule
\end{tabular}
\end{table}

In Table \ref{tab:sim1} we present the results of 20,000 simulations from the above setting
with a sample size of \(n=500\), \(n=1,000\), \(n=2,000\), and \(n=4,000\) subjects. We estimate in each
simulation the parameters
\begin{align*}
  &\estt{\psi} =    %% \estt{\theta}^{(0)}-\estt{\theta}^{(1)}=\pr(\potential{T^{\ast}}{0}\leq\tau,\:\potential{\epsilon^{\ast}}{0}=1)-\pr(\potential{T^{\ast}}{1}\leq\tau,\:\potential{\epsilon^{\ast}}{1}=1) \\
  \estt{\theta}^{(1)}-\estt{\theta}^{(0)}=\pr(\potential{T^{\ast}}{1}\geq\tau)-\pr(\potential{T^{\ast}}{0}\geq\tau) \\
  &\esty{\psi} = \esty{\theta}^{(1)}-\esty{\theta}^{(0)}=\E[\potential{Y}{1} \mid \potential{T^{\ast}}{1}\geq\tau]-\E[\potential{Y}{0} \mid \potential{T^{\ast}}{0}\geq\tau]
\end{align*}
based on the estimator \(\esty{\widetilde{\psi}}\) that ignores baseline
covariate information \eqref{eq:ipw}, and the one-step estimator,
\(\esty{\widehat{\psi}}\), derived from the efficient influence function \eqref{eq:eiffull}.
The nuisance models for \(\E(Y\mid A, R = 1,X_{1},X_{2})\), \(\pr(R=1\mid A,X_1,X_2)\) are
based on a linear model and logistic model, respectively, with main effects of
$X_1$ and $X_{2}$ and stratified by treatment. 

Similarly, Kaplan-Meier estimators are used to obtain an initial estimator
\(\estt{\widetilde{\psi}}\) of the risk-difference. Subsequently, the efficient
one-step estimator \(\estt{\widehat{\psi}}\) is derived based on the EIF
% \eqref{eq:teif},
where the nuisance model for the hazard of a terminal event is a Cox regression
with main effects $X_1$ and $X_{2}$ and baseline hazard stratified by treatment.
The censoring distribution is estimated using a Kaplan-Meier estimate separately
in each treatment arm.

The true parameter values are calculated numerically by Monte Carlo integration
from a large ($n=10^{8}$) simulated data set without censoring or missing data.
Resulting values were \(\esty{\psi}=2.790\) and \(\estt{\psi}=0.0259\).

\begin{table}[htbp]
  \caption{Simulation results based on 20,000 replications in the scenario with
    parameters defined in Table \ref{tab:simpar}.\label{tab:sim1}}
  \centering
\begin{tabular}{lrrrrrrr}
  \toprule
  \addlinespace
  \multicolumn{8}{c}{$n=500$} \\
  \addlinespace
  & Mean & Bias & SE & SD & SE/SD & Coverage & Rel.eff\\
  \midrule
Naive ($\esty{\widetilde{\psi}}$) & 2.8093 & 0.0191 & 1.7020 & 1.7049 & 0.9983 & 0.9476 & 1.0000\\
Adjusted ($\esty{\widehat{\psi}}$) & 2.7987 & 0.0086 & 1.2198 & 1.2273 & 0.9939 & 0.9494 & 0.7199\\
Naive ($\estt{\widetilde{\psi}}$) & 0.0260 & 0.0001 & 0.0280 & 0.0283 & 0.9900 & 0.9470 & 1.0000\\
Adjusted ($\estt{\widehat{\psi}}$) & 0.0260 & 0.0001 & 0.0279 & 0.0281 & 0.9926 & 0.9486 & 0.9941\\
  \bottomrule
  \addlinespace
  \multicolumn{8}{c}{$n=1,000$} \\
  \addlinespace
  & Mean & Bias & SE & SD & SE/SD & Coverage & Rel.eff\\
  \midrule
Naive ($\esty{\widetilde{\psi}}$) & 2.7919 & 0.0017 & 1.2046 & 1.2030 & 1.0013 & 0.9502 & 1.0000\\
Adjusted ($\esty{\widehat{\psi}}$) & 2.7814 & -0.0088 & 0.8643 & 0.8705 & 0.9929 & 0.9490 & 0.7236\\
Naive ($\estt{\widetilde{\psi}}$) & 0.0257 & -0.0002 & 0.0199 & 0.0199 & 0.9994 & 0.9504 & 1.0000\\
Adjusted ($\estt{\widehat{\psi}}$) & 0.0257 & -0.0002 & 0.0198 & 0.0198 & 0.9996 & 0.9511 & 0.9944\\
  \bottomrule
  \addlinespace
  \multicolumn{8}{c}{$n=2,000$} \\
  \addlinespace
  & Mean & Bias & SE & SD & SE/SD & Coverage & Rel.eff\\
  \midrule
Naive ($\esty{\widetilde{\psi}}$) & 2.7761 & -0.0141 & 0.8521 & 0.8581 & 0.9929 & 0.9474 & 1.0000\\
Adjusted ($\esty{\widehat{\psi}}$) & 2.7826 & -0.0075 & 0.6115 & 0.6131 & 0.9974 & 0.9498 & 0.7145\\
Naive ($\estt{\widetilde{\psi}}$) & 0.0258 & -0.0001 & 0.0141 & 0.0141 & 0.9977 & 0.9502 & 1.0000\\
Adjusted ($\estt{\widehat{\psi}}$) & 0.0258 & -0.0001 & 0.0140 & 0.0140 & 0.9991 & 0.9498 & 0.9923\\
  \bottomrule
  \addlinespace
  \multicolumn{8}{c}{$n=4,000$} \\
  \addlinespace
  & Mean & Bias & SE & SD & SE/SD & Coverage & Rel.eff\\
  \midrule
Naive ($\esty{\widetilde{\psi}}$) & 2.7859 & -0.0043 & 0.6027 & 0.6028 & 0.9998 & 0.9478 & 1.0000\\
Adjusted ($\esty{\widehat{\psi}}$) & 2.7860 & -0.0041 & 0.4326 & 0.4324 & 1.0006 & 0.9494 & 0.7173\\
Naive ($\estt{\widetilde{\psi}}$) & 0.0258 & -0.0001 & 0.0100 & 0.0101 & 0.9908 & 0.9484 & 1.0000\\
Adjusted ($\estt{\widehat{\psi}}$) & 0.0258 & -0.0001 & 0.0099 & 0.0100 & 0.9909 & 0.9476 & 0.9931\\
  \bottomrule
\end{tabular}
\end{table}

\begin{figure}[htbp]
  \centering
  \includegraphics{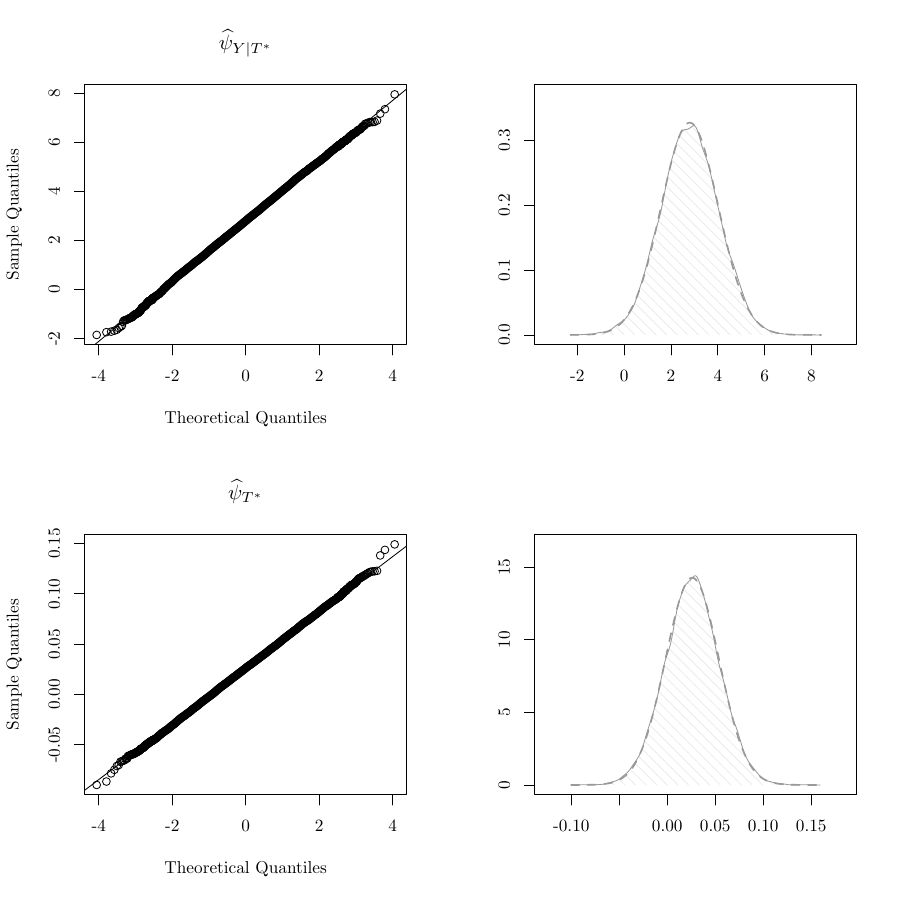}
  \caption{Normal approximation of the simulation study of
    the parameter estimates at $n=500$.}
  \label{fig:simres1}
\end{figure}

From Table \ref{tab:sim1} we confirm the consistency of both estimators and the estimates of the
asymptotic variance obtained from the variance of the respective influence
functions reflected in the nice agreement between the empirical average of the estimated
standard errors (SE) and the standard deviation of the parameter estimates over
the 20,000 simulation iterations (SD), as well as the estimated coverage of the
95\% Wald confidence limits. The Gaussian approximation is excellent already at $n=500$ (see Figure \ref{fig:simres1}).
Furthermore, as expected the one-step estimator
based on the efficient influence function is here considerably more efficient
for the parameter \(\esty{\psi}\)  (around 29\% smaller standard errors in the
covariate adjusted
estimator), whereas the efficiency gains are minor for \(\estt{\psi}\) (around
0.7\% smaller standard errors) due to
the weaker association between the covariates $X_{1}, X_{2}$ and the
time-to-event outcomes in this simulation. 

We next employ the proposed closed testing procedure as well as the Bonferroni-Holm procedure for testing $H_{Y|T^\ast}$ and $H_{T^\ast}$ to each simulated data set to assess their performance in terms of power. Results are summarized in Table \ref{powertab}.  

\begin{table}
  \small
  \centering
\caption{Power to reject either $H_{Y\mid T^{\ast}}$ or $H_{ T^{\ast}}$ or both hypotheses at a nominal significance level $\alpha=0.025$ and with superiority/non-inferiority margins $\esty{\delta}=\estt{\delta}=0$. \label{powertab} }
\begin{tabular}{lccccccc}
\toprule
\addlinespace
\multicolumn{2}{c}{}&\multicolumn{3}{c}{Proposed testing procedure}&\multicolumn{3}{c}{Bonferroni-Holm procedure}\\
\addlinespace
  & Sample size & $H_{Y\mid T^{\ast}}$ & $H_{T^{\ast}}$ & $H_{Y\mid T^{\ast}}$ and $H_{T^{\ast}}$ & $H_{Y\mid T^{\ast}}$ & $H_{T^{\ast}}$ & $H_{Y\mid T^{\ast}}$ and $H_{T^{\ast}}$\\
\midrule

Adjusted & 500 & 0.5660 & 0.1462 & 0.0868 & 0.5305 & 0.1231 & 0.0807\\
Naive & 500 & 0.3126 & 0.1338 & 0.0589 & 0.2875 & 0.1109 & 0.0538\\
Adjusted & 1000 & 0.8713 & 0.2535 & 0.2196 & 0.8471 & 0.2386 & 0.2145\\
Naive & 1000 & 0.5850 & 0.2426 & 0.1616 & 0.5479 & 0.2128 & 0.1525\\
Adjusted & 2000 & 0.9944 & 0.4529 & 0.4498 & 0.9917 & 0.4517 & 0.4492\\
Naive & 2000 & 0.8863 & 0.4465 & 0.4036 & 0.8627 & 0.4308 & 0.3963\\
Adjusted & 4000 & 1.0000 & 0.7388 & 0.7388 & 1.0000 & 0.7388 & 0.7388\\
Naive & 4000 & 0.9956 & 0.7327 & 0.7300 & 0.9944 & 0.7322 & 0.7297\\
\bottomrule
\end{tabular}
\end{table}

From Table \ref{powertab} we note a substantial power gain when comparing our proposed testing procedure based on the one-step estimators to the traditional Bonferroni-Holm procedure based on the unadjusted estimators. In particular, a substantial power gain is obtained by using the one-step estimators over the unadjusted estimators. A smaller but still appreciable gain in power is seen from using the proposed testing strategy instead of the Bonferroni-Holm procedure. 

To assess also type 1 error of the proposed testing procedure under the global null hypothesis $H_{Y\mid T^{\ast}}\cap H_{ T^{\ast}}$  we consider a simulation scenario where data in the active treatment arm ($A=1$) are generated according to the specification for the placebo arm ($A=0$) in Table \ref{tab:simpar}.  Again we simulate 20,000 data sets and summarize the performance of our proposed estimation and testing strategy in terms of type 1 error control in Table \ref{type1tab}.

\begin{table}
  \centering
\caption{Type 1 error for testing $H_{Y\mid T^{\ast}}\cap H_{ T^{\ast}}$, $H_{Y\mid T^{\ast}}$, and $H_{ T^{\ast}}$ under the global null using signed Wald tests  at a nominal significance level $\alpha=0.025$ and with superiority/non-inferiority margins $\esty{\delta}=\estt{\delta}=0$. \label{type1tab}  }
\begin{tabular}{lrrrr}
\toprule
  & Sample size & $H_{Y\mid T^{\ast}}\cap H_{ T^{\ast}}$ & $H_{Y\mid T^{\ast}}$ & $H_{ T^{\ast}}$ \\
\midrule

Adjusted& 500 & 0.0272 & 0.0249 & 0.0275\\
Naive& 500 & 0.0270 & 0.0271 & 0.0279\\
Adjusted & 1000 & 0.0253 & 0.0256 & 0.0242\\
Naive& 1000 & 0.0254 & 0.0256 & 0.0255\\
Adjusted& 2000 & 0.0240 & 0.0243 & 0.0245\\
Naive & 2000 & 0.0258 & 0.0251 & 0.0248\\
Adjusted & 4000 & 0.0246 & 0.0243 & 0.0249\\
Naive & 4000 & 0.0260 & 0.0250 & 0.0246\\

\bottomrule
\end{tabular}
\end{table}

From Table \ref{type1tab} we conclude that the type 1 error is controlled well at the nominal 2.5\% significance level in all scenarios and when testing both $H_{Y\mid T^{\ast}}\cap H_{ T^{\ast}}$, $H_{Y\mid T^{\ast}}$, and $H_{ T^{\ast}}$. 

To also explore the performance in a situation where there is a stronger association between 
covariates and the terminal event of interest, we consider a scenario with $n=2000$ identical to the parameters in
Table \ref{tab:simpar} except that for the cause-specfic hazard for first major irreversible kidney
event, \(\epsilon=1\), we increase the effect of the covariate $X_{1}$ to
$\beta_{\epsilon=1,1}^{(A=1)} = \beta_{\epsilon=1,1}^{(A=0)} = -0.15$. The summarized
results of 20,000 simulated data sets are shown in Table \ref{tab:sim2}.

%Finally, we also examined the performance in the null %model where
%all the regression parameters were set to be identical %in the two arms $A=0$ and
%$A=1$, with the parameters in the active arm $A=1$ being %set to the values of
%the corresponding parameters in the control arm $A=0$ %defined in Table
%\ref{tab:simpar}. The results are shown in Table %\ref{tab:sim2} and we see that
%both estimators control the type I error at the nominal %level.

\begin{table}[htbp]
    \caption{Simulation results based on 20,000 replications in a scenario with stronger covariate effect on the
    cause-specific hazard for the primary event.\label{tab:sim2}}
  \centering
\begin{tabular}{lrrrrrrr}
  \toprule
  \addlinespace
  & Mean & Bias & SE & SD & SE/SD & Coverage & Rel.eff\\
  \midrule
Naive ($\esty{\widetilde{\psi}}$) & 2.1252 & -0.0152 & 0.8388 & 0.8407 & 0.9977 & 0.9493 & 1.0000\\
Adjusted ($\esty{\widehat{\psi}}$) & 2.1335 & -0.0069 & 0.6713 & 0.6725 & 0.9982 & 0.9508 & 0.7999\\
Naive ($\estt{\widetilde{\psi}}$) & 0.0350 & -0.0001 & 0.0192 & 0.0194 & 0.9899 & 0.9491 & 1.0000\\
Adjusted ($\estt{\widehat{\psi}}$) & 0.0350 & -0.0001 & 0.0144 & 0.0144 & 0.9948 & 0.9487 & 0.7433\\
  \bottomrule
\end{tabular}
\end{table}

From Table \ref{tab:sim2} we note that efficiency gains for both estimators are now
substantial with approximately 26\% reduction in standard errors of the
efficient estimator $\estt{\widehat{\psi}}$ compared to the Kaplan-Meier. This
simulation demonstrates that the efficiency gains in a realistic setting can be
considerable for both target parameters.

In a final simulation study, we investigate robustness of the proposed estimators under  mis-specification of the nuisance models and substantial censoring. 
We consider a scenario identical to the parameters in
Table \ref{tab:simpar} except that censoring times are now generated according to a uniform distribution from 0 to 4 years. This results in 48\% censoring.  

The nuisance model for $\E(Y| A, R=1, X_{1}, X_{2})$ is mis-specified by not including treatment specific effects of $X_{2}$ and $X_{2}$. For the nuisance model $\pr(R=1| A, X_1, X_2)$ the covariates $X_1$, $X_2$ are removed entirely from the logistic regression. Similarly the nuisance model to obtain the one step estimator of $\estt{\widehat{\psi}}$ is mis-specified as a Cox regression including only a treatment effect.

In table \ref{tab:simrob} we present results from 20000 simulations according to the above setting for each of the sample sizes  \(n=500\), \(n=1,000\), \(n=2,000\), and \(n=4,000\) subjects. 

\begin{table}[htbp]
    \caption{Simulation results based on 20,000 replications in a scenario with
      substantial censoring and mis-specified nuisance models. \label{tab:simrob}}
    \centering
\begin{tabular}{lrrrrrrr}
\toprule
  \addlinespace
  \multicolumn{8}{c}{$n=500$} \\
  \addlinespace
& Mean & Bias & SE & SD & SE/SD & Coverage & Rel.eff\\
\midrule
Naive ($\esty{\widetilde{\psi}}$) & 2.8077 & 0.0092 & 1.7418 & 1.7439 & 0.9988 & 0.9462 & 1.0000\\
Adjusted ($\esty{\widehat{\psi}}$) & 2.7988 & 0.0003 & 1.2793 & 1.2829 & 0.9972 & 0.9502 & 0.7357\\
Naive ($\estt{\widetilde{\psi}}$) & 0.0259 & 0.0000 & 0.0342 & 0.0347 & 0.9877 & 0.9473 & 1.0000\\
Adjusted ($\estt{\widehat{\psi}}$) & 0.0259 & 0.0000 & 0.0344 & 0.0347 & 0.9909 & 0.9480 & 1.0026\\
 \bottomrule
  \addlinespace
  \multicolumn{8}{c}{$n=1,000$} \\
  \addlinespace

Naive ($\esty{\widetilde{\psi}}$) & 2.7900 & -0.0085 & 1.2329 & 1.2299 & 1.0024 & 0.9492 & 1.0000\\
Adjusted ($\esty{\widehat{\psi}}$) & 2.7797 & -0.0188 & 0.9059 & 0.9089 & 0.9968 & 0.9504 & 0.7390\\
Naive ($\estt{\widetilde{\psi}}$) & 0.0257 & -0.0002 & 0.0243 & 0.0244 & 0.9967 & 0.9492 & 1.0000\\
Adjusted ($\estt{\widehat{\psi}}$) & 0.0257 & -0.0002 & 0.0244 & 0.0244 & 0.9983 & 0.9496 & 1.0013\\
 \bottomrule
  \addlinespace
  \multicolumn{8}{c}{$n=2,000$} \\
  \addlinespace
Naive ($\esty{\widetilde{\psi}}$) & 2.7771 & -0.0214 & 0.8721 & 0.8770 & 0.9945 & 0.9468 & 1.0000\\
Adjusted ($\esty{\widehat{\psi}}$) & 2.7833 & -0.0152 & 0.6409 & 0.6403 & 1.0010 & 0.9500 & 0.7301\\
Naive ($\estt{\widetilde{\psi}}$) & 0.0257 & -0.0002 & 0.0172 & 0.0173 & 0.9970 & 0.9506 & 1.0000\\
Adjusted ($\estt{\widehat{\psi}}$) & 0.0257 & -0.0002 & 0.0173 & 0.0173 & 0.9978 & 0.9507 & 1.0006\\
 \bottomrule
  \addlinespace
  \multicolumn{8}{c}{$n=4,000$} \\
  \addlinespace
Naive ($\esty{\widetilde{\psi}}$) & 2.7856 & -0.0129 & 0.6168 & 0.6179 & 0.9983 & 0.9476 & 1.0000\\
Adjusted ($\esty{\widehat{\psi}}$) & 2.7856 & -0.0129 & 0.4534 & 0.4545 & 0.9976 & 0.9488 & 0.7355\\
Naive ($\estt{\widetilde{\psi}}$) & 0.0258 & -0.0001 & 0.0122 & 0.0123 & 0.9916 & 0.9488 & 1.0000\\
Adjusted ($\estt{\widehat{\psi}}$) & 0.0258 & -0.0001 & 0.0122 & 0.0123 & 0.9920 & 0.9492 & 1.0003\\

\bottomrule
\end{tabular}
\end{table}

From Table \ref{tab:simrob} we note some loss of precision in both the naive and adjusted estimators when compared to Table \ref{tab:sim1} summarizing our first simulation study.  This is to be expected due to more extreme censoring. When it comes to the impact of misspecifying the nuisance models, however, we see no indication that this affects estimator performance negatively. The adjusted estimators are virtually unbiased, have good coverage, and similar efficiency gains are observed as in the first simulation study.

\section{Application}\label{sec:application}

The FLOW (Evaluate Renal Function with Semaglutide Once Weekly) clinical kidney outcome trial randomised 3,533 patients 1:1 to receive either placebo or semaglutide on top of standard of care \citep{Flow2024}. Semaglutide is a glucagon-like peptide-1 receptor agonist (GLP-1 RA) approved for treatment of type 2 diabetes. All patients had type 2 diabetes and had high-risk chronic kidney disease. High risk kidney disease patients were selected according to the estimated glomerular filtration rate (eGFR) per serum creatinine and urinary albumin to creatinine ratio (UACR). The trial duration was 5 years with a median follow-up time of 3.4 years. The trial objective was to demonstrate that semaglutide delayed the progression of kidney impairment and lowered the risk of kidney and cardiovascular mortality compared to placebo, both added to standard-of-care, in subjects with type 2 diabetes and chronic kidney disease \citep{Flow2024}. The primary endpoint was time to first composite major kidney disease event consisting of; a sustained decline in eGFR above 50 $\%$ relative to baseline, sustained eGFR < 15 $\text{mL/min/1.73m}^2$, renal replacement therapy (dialysis or transplantation), renal or cardiovascular death. The annual rate of change in eGFR from randomisation, total eGFR slope, was a confirmatory secondary endpoint. The trial was event driven and employed a group sequential design with a planned interim for efficacy after two thirds of the primary endpoint events had occurred. The trial was stopped at interim following the interim evaluation.

For this application, the eGFR measurement at landmark year 2 after randomization will constitute the surrogate marker. A higher eGFR is indicative of a better renal function with an eGFR of more than $\text{90 mL/min/1.73m}^2$ indicating a normal or high kidney function (\cite{stevens2024kdigo}). Thus, $\esty{\psi}= \E[\potential{Y}{1} \mid \potential{T^{\ast}}{1}\geq\tau]-\E[\potential{Y}{0} \mid \potential{T^{\ast}}{0}\geq\tau] > 0$ corresponds to a better average renal function after two years on semaglutide treatment without terminal events when compared to average renal function after two years on placebo treatment without terminal events. Accordingly, as one of two parts of the joint assessment of treatment effect we consider the null-hypothesis:

$$\esty{H}:\: \esty{\psi} \leq 0.$$ 

Moreover, time to first major kidney disease event or death from other causes define the onset of terminal event. A lower risk of having a terminal event two years after randomization corresponds to a beneficial effect of treatment. Thus, $\estt{\psi}=\pr(\potential{T^{\ast}}{1}\geq\tau)-\pr(\potential{T^{\ast}}{0}\geq\tau) >0$ corresponds to a beneficial effect of semaglutide on the risk of having a terminal event.  Therefore, as the remaining part of our joint assessment of treatment effect, we test the null-hypothesis:

$$\estt{H}:\: \estt{\psi} \leq 0.$$

\noindent We estimate $\esty{\psi}$ and $\estt{\psi}$ using the developed methodology and based on the same nuisance models that we applied in the simulation study. Next we test the hypotheses $\esty{H}$ and $\estt{H}$ using the proposed closed testing strategy. The results of this analysis of the FLOW data are presented in Table~\ref{tab:flowres}.

\begin{table}[!htbp]
  \centering
  \caption{Analysis results based on FLOW trial data.\label{tab:flowres}}
  \centering
  \begin{tabular}{lcccc}
  \toprule
  \addlinespace
  \multicolumn{5}{c}{\textbf{On surrogate marker in the absence of terminal event, eGFR at year 2 $(\tau=2)$}} \\
  \addlinespace
    & & Estimate %& Standard error
    & 95 $\%$ CI & P-value \\
  \midrule
  Placebo: & $\esty{\theta}^{(0)} = \E[\potential{Y}{0} \mid \potential{T^{\ast}}{0}\geq\tau]$ &
                                                                                41.541 &
                                                                                % 0.412 &
[40.735 ; 42.348] & - \\
  Sema: &$\esty{\theta}^{(1)} = \E[\potential{Y}{1} \mid \potential{T^{\ast}}{1}\geq\tau]$ &
                                                                            44.222 &
                                                                            % 0.410 &
[43.418 ; 45.026] & - \\
    Sema - Placebo: &$\esty{\psi} = \esty{\theta}^{(1)}-\esty{\theta}^{(0)}$ & 2.681 &
    % 0.489 &
[1.721 ; 3.640] & $<0.0001$ \\
  \bottomrule
  \addlinespace
  \multicolumn{5}{c}{\textbf{On terminal event, major kidney disease events or death}} \\
  \addlinespace
  & & Estimate
    % & Standard error
    & 95 $\%$ CI & P-value\\
  \midrule
   Placebo: &
              $\estt{\theta}^{(0)} = \pr(\potential{T^{\ast}}{0}\geq\tau)$ & 0.8697 &
%0.0070 &
[0.8540 ; 0.8855] & -\\
   Sema: &
           $\estt{\theta}^{(1)} = \pr(\potential{T^{\ast}}{1}\geq\tau)$ & 0.9012
               % & 0.0059
  & [0.8873 ; 0.9152] & -\\
    Placebo - Sema: &$\estt{\psi} = \estt{\theta}^{(1)}-\estt{\theta}^{(0)}$ & 0.0315 &
    % 0.0091 &
  [0.0106 ; 0.0524] & $0.0032$\\
  \bottomrule
  \addlinespace
  \multicolumn{5}{c}{\textbf{One-sided tests: Signed Wald test}} \\
  \addlinespace
  &  Hypothesis  & Test-statistic & P-value\\
  \midrule
Sema - Placebo: & $\esty{H}: \esty{\psi}\leq 0$ & 29.996 & $<0.0001$ \\
Placebo - Sema: & $\estt{H}: \estt{\psi} \leq 0$ & 8.697 & 0.0016 \\
  \bottomrule
    \addlinespace
  \multicolumn{5}{c}{\textbf{Intersection test: Signed Wald intersection test}} \\
  \addlinespace
  &  Hypothesis  & Test-statistic & P-value\\
  \midrule
Sema vs. Placebo: & $H_{Y\mid T^{\ast}}\cap H_{ T^{\ast}}$ & 41.716 & $<0.0001$  \\
  \bottomrule
\end{tabular}
\end{table}

\noindent From Table~\ref{tab:flowres} we conclude that there is evidence of a clear benefit of semaglutide in lowering the risk of terminal events after two years of treatment. Compared to placebo there is also evidence of a clear improvement of kidney function in terms of increased eGFR after two years of treatment with semaglutide among those that are still alive and without major kidney events. 

The naive method estimates $\esty{\psi}$ to $2.592$ which is similar to the adjusted estimate in Table~\ref{tab:flowres}. However, the resulting 95\% CI is $[1.298 ; 3.886]$ which is substantially wider than the 95\% CI
presented in Table~\ref{tab:flowres}. This reflects that the standard error decreases from $0.660$ for the naive method to $0.489$ with the proposed adjustment. The naive estimate for $\estt{\psi}$ is $0.0304$ with 95\% CI $[ 0.00938; 0.0515]$ and, comparing to Table~\ref{tab:flowres}, adjustment offers no significant precision gain in this case. We note that these observations reflect the findings of our simulation study well.

\section{Discussion}\label{sec:discussion}

Current practice to analyse decline in eGFR involves very explicit modelling of
eGFR profiles by means of random slope models \citep{Vonesh2019}. Such
simplifications may be hard to justify in studies such as the FLOW study. This
may lead to inadequate description of the actual behavior and  consequent loss
of power to detect a relevant decline in eGFR \citep{DeVries20204}. Moreover,
effects reported from these models are based on extrapolation beyond terminal
events and thus consider the impact of treatment in a hypothetical scenario
where terminal events can be prevented \citep{Kahan2020}. Finally, the random
slope models that are used  require intensive sampling of eGFR and as such pose
a burden for both study sponsors and study participants. In this paper we have
offered an alternative approach to jointly analyse eGFR and occurrence of terminal events that does not require such
strict assumptions and we have shown by simulation and example that this approach is attractive in terms of performance and precision. Our proposal also offers a clinically relevant and transparent evaluation of important aspects of risk/benefit in the context of terminal events.

While the proposed framework offers a robust and assumption-lean approach for
analyzing outcomes that are truncated by a terminal event, we acknowledge that
the choice of the landmark time is a critical parameter that defines the
specific identified clinical functions being evaluated. In our analysis of the
FLOW trial, $\tau=2$ years was selected to align with established clinical
milestones for assessing kidney function. However, as different landmark times
may represent different clinical insights, future applications should explore
the robustness of treatment effects across multiple clinically relevant time
points to ensure stable clinical conclusions. Furthermore, although we have
extended the methodology to accommodate covariate-dependent missingness, the
framework still relies on certain conditional independence assumption for the
missing data and censoring mechanisms. Building upon the Efficient Influence Functions
(EIF) derived in this work, future research could develop formal sensitivity
analyses, such as delta-adjustment, to quantify how potential departures
from these assumptions might impact the reported results. Such extensions would
further enhance the utility of our semi-parametric approach in settings where
latent health status might introduce informative missingness or censoring.

In our exposition we focused on a formalized assessment of treatment effects on
one clinical score and any terminal event.
However, the estimation procedure is easily extended to handle estimation of
more clinical scores and specific types of terminal events in a competing risk
scenario.  This extension would facilitate a more detailed evaluation of treatment impact by disentangling primary clinical events, like major kidney failure, from unrelated terminal events such as cardiovascular death.
To also extend the closed testing procedure we would need to consider a generalized version of the signed Wald test (\ref{SignWald}) for the intersection hypotheses. Specifically, in our scenario we may rewrite (\ref{SignWald}) as:
$$
\inf_{u\in W_{1}\cap W_{2}}\|\hat{u}-u\|^{2},
$$
where $W_{j}=\{u:\: \sqrt{\Sigma}_{j}u\leq0\},\:\: j=1,2$ denote the half-spaces encoded by the constraint $\sqrt{\Sigma}u\leq 0$. With this rewrite it is easy to express the signed Wald test for the intersection of  multiple superiority/non-inferiority hypotheses \{$H_{l}\}_{l=1,\ldots, L}$  as:
\begin{equation}\label{signWald-ext}
SW_{n,\cap_{l=1}^{L}H_{l}}=\inf_{u\in \cap_{j=1}^{J}W_{j}}\|\hat{u}-u\|^{2},
\end{equation}
where again $W_{j}=\{u:\: \sqrt{\Sigma}_{j}u\leq0\},\:\: j=1,\ldots, J$ denote the half-spaces encoded by the constraint $\sqrt{\Sigma}u\leq 0$.

There is no closed form expression to calculate the $SW_{n,\cap_{l=1}^{L}H_{l}}$ in general. However, since the right hand side of (\ref{signWald-ext})  is identified as the minimal distance from a point to an intersection of half-spaces it can be computed numerically by Dykstras projection algorithm \citep{Dykstra1983}. This effectively means that we can simulate the null-distribution of the signed Wald test for all intersection hypotheses needed to  enable a generalized closed testing procedure. Specifically we can simulate the null distribution by repeatedly simulating zero mean standard normal variables $U_{i}$ and calculating their distance to the intersection of half-spaces. We plan to investigate this proposal in more detail in future research with the following two applications in mind. 

Firstly, from a FLOW perspective, such an extension would facilitate that we could include additional surrogate markers such as the urin albumin-creatinin ratio. We could also provide a more detailed evaluation of the impact of treatment specifically on major kidney events as well as death from other causes. 

Secondly, if, in the FLOW application,  we had only rejected one of the hypotheses $\esty{H}$ and $\estt{H}$, an overall conclusion about treatment benefit would be difficult to make based on this evidence alone. To mitigate this situation a utility assessment of overall benefit can be added by also testing the null-hypothesis:
$$
H_{Y^{\ast}| T^{\ast},T^{\ast}}: \E(U^{(1)})-\E(U^{(0)})\leq 0,
$$
where $U^{(a)}=Y^{(a)}\cdot I(T^{\ast(a)}\geq\tau)+\Gamma\cdot I(T^{\ast(a)}<\tau)$ for some unfavorable value $\Gamma$.

In order to apply the above extension of the signed Wald test to the hypotheses
$\esty{H}$, $\estt{H}$, $H_{Y^{\ast}| T^{\ast},T^{\ast}}$, and intersections thereof we
need to produce a consistent linear asymptotically normal estimator of
$\E(U^{(1)})-\E(U^{(0)})$ and identify its influence function. However, the
quantities
$\E(U^{(a)})=\esty{\theta}^{(a)}\cdot\pr(\potential{T^{\ast}}{a}\geq\tau)+\Gamma\cdot \pr(\potential{T^{\ast}}{a}<\tau)$
are easily estimated by plugging in the estimates of $\esty{\theta}^{(a)}$ and
$\pr(\potential{T^{\ast}}{a}\geq\tau)$ that were derived in Section \ref{sec:est}. The influence
function of the resulting plugin estimator can be derived by standard arguments.

As a cautionary remark, we also want to point out that in our framework change
from baseline in clinical scores and actual clinical score values at a landmark
time can not be used interchangeably. For instance, in FLOW, a baseline
measurement $X_{1}$ of the eGFR score is available. It would therefore be
natural to move from assessing treatment effect on the eGFR score $Y$ at a
landmark time to use $\tilde{Y}=Y-X_{1}$ for that assessment. Note however that
in our setup this would lead to contrasting
\begin{align*}
  \psi_{\tilde{Y}| T^{\ast}}&=\E[\potential{\tilde{Y}}{1} \mid \potential{T^{\ast}}{1}\geq\tau]-\E[\potential{Y}{0} \mid \potential{T^{\ast}}{0}\geq\tau] \\
  &=\psi_{Y| T^{\ast}}-\{\E[X_{1}\mid T^{\ast}\geq\tau,\: A=1]- \E[X_{1}\mid T^{\ast}\geq\tau,\: A=0]\}.
\end{align*}
Since the last term on the right hand side above is not guaranteed to be zero unless $X_1$ is independent of $I(T^{\ast}\geq\tau)$ given $A$ we are effectively targeting another parameter to assess effect. This means that estimated treatment effects based on either $Y$ or $\tilde{Y}$ are not comparable due to the selection process instated by truncation.

%We note that an alternative approach could be to change the target parameter to
%a single utility measure. Formally, a utility assessment of benefit could be
%considered by testing the null-hypothesis:
%$$
%H_{Y^{\ast}| T^{\ast},T^{\ast}}: \E(U^{(1)})-\E(U^{(0)})\leq 0,
%$$
%where $U^{(a)}=Y^{(a)}\cdot I(T^{\ast(a)}>\tau)+\Gamma\cdot %I(T^{\ast(a)}\leq\tau)$ for some unfavorable value $\Gamma$ on the clinical %measurement scale. Note that this add-on is particularly useful in scenarios where %both of the hypotheses $\esty{H}$ and $\estt{H}$ are not formally rejected.
%To test $H_{Y^{\ast}| T^{\ast},T^{\ast}}$, a signed Wald test could be based on %the standardized estimate of $\E(U^{(1)})-\E(U^{(0)})$. The quantities %$\E(U^{(a)})=\esty{\theta}^{(a)}\cdot\pr(\potential{T^{\ast}}{a}>\tau)+\Gamma\cdot %\pr(\potential{T^{\ast}}{a}\leq\tau)$ are estimated by plugging in estimates of %$\esty{\theta}^{(a)}$ and $\pr(\potential{T^{\ast}}{a}\leq\tau)$. An efficient one-%step estimator of $\esty{\theta}^{(a)}$ is already derived in Section %\ref{sec:est} and an efficient one-step estimator of $\pr(\potential{T^{\ast}}%{a}\leq\tau)$ can be derived in similar vein as the efficient one-step estimator for $\estt{\theta}^{(a)}$. Finally the stacking method as described in Section \ref{sec:est} can be applied in combination with the Delta Theorem to obtain a consistent standard error of the resulting estimator of $\E(U^{(1)})-\E(U^{(0)})$.

Finally, we would like to emphasize that the developed methodology has potential to be used in many other disease areas besides chronic kidney disease. Examples of other areas where we see a potential for this methodology include KCCQ scores in heart failure patients \citep{QCCQ} and MoCA scores in dementia patients \citep{MOCA}.

%%%%%%%%%%%%%%%%%%%%%%%%%%%%%%%%%%%%%%%%%%%%%%%%%%

\bibliographystyle{apalike}
\bibliography{ref}

\appendix
\clearpage
\appendix
%%%%%%%%%%%%%%%%%%%%%%%%    %%%%%%%
%% Influence functions
%%%%%%%%%%%%%%%%%%%%%%%%%%%%%%%
\section{Deriving the Randomization Augmented Influence Function for \(\esty{\theta}^{(a)}\)}\label{sec:eif:thetay}

We first describe the subtangent space directly associated with the randomization
assumption \ref{as:randomization}. By a orthogonality argument, the projection of
any influence function onto this subtangent space will yield a reduced or
equivalent variance. We then calculate the projection of the influence function
associated with the plug-in estimator of $\esty{\theta}^{(a)}$ to obtain what we denote as
the randomization augmented influence function.

Due to the treatment
randomization assumption \ref{as:randomization}, the log-likelihood for the
observed data, \(\{X,A,T, \Delta(\tau),R, RY\}\), has the following decomposition
\begin{align*}
\log\{f(T, \Delta(\tau),R, RY\mid A, X)\} + \log\{f(A)\} + \log\{f(X)\}.
\end{align*}
It follows that the tangent space as a subspace of the Hilbert space $\mathcal{H}$ of
$L^2_{P_0}$ zero mean functions endowed with covariance inner product is given
by
\begin{align*}
  \mathcal{T}_{1} \oplus
  \mathcal{T}_{2} \oplus
  \mathcal{T}_{3},
\end{align*}
where
\begin{align*}
  &\mathcal{T}_{1} \subset \{ h(X, A, T, \Delta(\tau),R, RY) \in \mathcal{H} \mid \E[h(X, A,T, \Delta(\tau),R, RY)\mid A,X] = 0\}, \\
  &\mathcal{T}_{2} = \{ h(A) \in \mathcal{H} \mid \E[h(A)] = 0\}, \\
  &\mathcal{T}_{3} = \{ h(X) \in \mathcal{H} \mid \E[h(X)] = 0\},
\end{align*}
We are interested in the orthogonal complement to $\mathcal{T}_{2}$ and
$\mathcal{T}_{3}$ contained in
\begin{align*}
\{ h(X,A) \in \mathcal{H} \mid \E[h(X,A)] = 0\}.
\end{align*}
Along the lines of \cite{Zhang_2008}, we get that
\begin{align*}
  \mathcal{T}^{\perp} =\{ h(X,A) \in \mathcal{H} \mid \E[h(A,X)] = 0\} \cap \mathcal{T}_{2}^{\perp} \cap \mathcal{T}_{3}^{\perp} = \{ h(X,A) \in \mathcal{H} \mid \E[h(X,A)\mid X] = 0\}.
\end{align*}
As \(A\) is binary with \(\pi_0(a) = \pr(A=a)\), we see that
\begin{align}\label{eq:compltangent}
  \mathcal{T}^{\perp} = \{ (A-\pi(1))h(X) \mid \E[h(X)^{2}] <\infty \}.
\end{align}
A consistent estimator for \(\esty{\theta}^{(a)}(P_0)\) is immediately obtained
from the plugin (inverse probability weighting) estimator
\begin{align}\label{eq:ipw}
  \esty{\widetilde{\theta}}^{(a)} = \Pn \frac{I(R=1, A=a)}
  {\pr_{n}I(R=1, A=a)} Y,
\end{align}
which has influence function
\begin{align*}
  \esty{\widetilde{\phi}}^{(a)}(Z; P) = \frac{I(A=a)R}{\pr_P(A=a)\pr_P(R=1\mid A=a)}\{Y - \esty{\theta}^{(a)}(P)\}.
\end{align*}
The randomization augmented influence function for $\esty{\theta}^{(a)}(P)$ is now derived as
\begin{align*}
  \esty{\phi}^{(a)}(Z; P) = \esty{\widetilde{\phi}}^{(a)}(Z; P) - \Pi\left(\esty{\widetilde{\phi}}^{(a)}(Z; P) \mid \mathcal{T}^{\perp}\right)
\end{align*}
The projection term is calculated as follows. An element in
\(\mathcal{T}^{\perp}\) has the form \((A-\pi(1))h(X)\) for an arbitrary
element \(h\). We need to find \(h^{\ast}\) such that
\(\esty{\widetilde{\phi}}^{(a)}(Z; P) - (A-\pi(1))h^{\ast}(X)\) is orthogonal to all of
\(\mathcal{T}^{\perp}\), that is,
\begin{align*}
  \forall h\colon
  \E_P\left(\left\{
  \frac{I(A=a)R}{\pi(a)\rho(a)}\{Y - \esty{\theta}^{(a)}(P)\} - (A-\pi(1))h^{\ast}(X)
  \right\}(A-\pi(1))h(X)
  \right) = 0,
\end{align*}
from which it follows that
\begin{align*}
  \E_P\left(\left\{
  \frac{I(A=a)R}{\pi(a)\rho(a)}\{Y - \esty{\theta}^{(a)}(P)\} - (A-\pi(1))h^{\ast}(X)
  \right\}(A-\pi(1)) \Big\vert X
  \right) = 0.
\end{align*}
This implies that
\begin{align*}
  h^{\ast}(X)(1-\pi(1))\pi(1)
  &= \frac{a-\pi(1)}{\pi(a)}\frac{1}{\rho(a)}\E_P\Big[I(A=a)R\,\E_P\left\{Y-\esty{\theta}^{(a)}(P)\big\vert X, A,R\right\} \Big\vert X\Big] \\
  & =
      \frac{a-\pi(1)}{\pi(a)}\frac{1}{\rho(a)}\pr_P\left(A=a, R=1 \mid X\right)\E_P\left\{Y-\esty{\theta}^{(a)}(P)\mid X, A=a, R=1\right\}
       \\ %% \pr[A=0, T^{\ast}>\tau \mid \mid X]
  & \overset{\ref{as:randomization}}{=}
      (a-\pi_{1})\frac{\rho(X,a)}{\rho(a)}\left\{Q(X,a)-\esty{\theta}^{(a)}(P)\right\},
\end{align*}
with \(Q(X,a; P) = \E_{P}\{Y\mid X, A=a, R=1\}\), and
\(\rho(X,a; P) =  \pr_P(R=1 \mid X, A=a)\).
It follows that
\begin{align}
  \begin{split}
  \esty{\phi}^{(a)}(Z; P)
  &= \frac{I(A=a)I(R=1)}{\pi(a)\rho(a)}\left\{Y - \esty{\theta}^{(a)}(P)\right\} \\
  &\qquad + \frac{\pi(a)-I(A=a)}{\pi(a)} \frac{\rho(X,a)}{\rho(a)}\left\{Q(X,a)-\esty{\theta}^{(a)}(P)\right\},
  \end{split}
\end{align}
because
\begin{align*}
\frac{(A-\pi(1))(a-\pi(1))}{\pi(1)(1-\pi(1))} = \frac{I(A=a)-\pi(a)}{\pi(a)}.
\end{align*}

\vspace*{1em}

\section{Asymptotic properties}\label{sec:app:asymp}

In the following we use the notation $\Pn$ to denote the empirical mean such
that $\Pn \widehat{Q}(Z) = n^{-1}\sum_{i=1}^n \widehat{Q}$ over the
i.i.d. observed data $Z_{1},\ldots,Z_{n}$ but keeping $\widehat{\mathcal{Q}}$
fixed. Similarly we use $\PE$ to denote the mean with respect to the true
data-generating process
$\PE \widehat{Q}(Z) = \int Q(z)\,dP(z)$.

Let \(\widehat{Q}(X,a)\) and \(\widehat{\rho}(X,a)\) be the
two misspecified regression models that converges to
\(Q^{\ast}(X,a)\not=Q(X,a; P)\) and \(\rho^{\ast}(X,a)\not=\rho(X,a;P)\) in the sense that
$\pr\big\{(Q^{\ast}(X,a)- \widehat{Q}(X,a))^2\big\}$ and
$\pr\big\{(\rho^{\ast}(X,a)- \widehat{\rho}(X,a))^2\big\}$ converges to zero in probability.
It follows that the estimating equation derived from the IF is still
consistent
\begin{align*}
  \E[\esty{\phi}^{(a)}(Z; \mathcal{Q}^{\ast})] &= 0 -
                                           \E\left[\frac{\{\pi(a)-I(A=a)\}}{\pi(a)}
                                           \frac{\rho(X, a)}{\rho(a)}
  \{
  Q^{\ast}(X,a) - \esty{\theta}^{(a)}
  \}\right] \\
  &= \E \left\{\frac{\E\left[\pi(a)-I(A=a)|X\right]}{\pi(a)}
    \frac{\rho(X,a)}{\rho(a)}\{Q^{\ast}(X,a) - \esty{\theta}^{(a)}\}
  \right\} = 0.
\end{align*}
where $\mathcal{Q}^{\ast} := \{Q^{\ast}, \rho^{\ast},
\pi(a), \rho(a), \esty{\theta}^{(a)} \mid a=0,1\}$. We can now decompose the one-step estimator in the following way. Define the
remainder term
$R(\widehat{\mathcal{Q}}) = \PE\esty{\phi}^{(a)}(Z; \widehat{\mathcal{Q}}) + \esty{\widetilde{\theta}}^{(a)} - \esty{\theta}^{(a)}$,
then direct calculations yield the following von-Mises expansion
\begin{align*}
  \esty{\widehat{\theta}}^{(a)} - \esty{\theta}^{(a)}
  &= \Pn \esty{\phi}^{(a)}(Z; \widehat{\mathcal{Q}}) + \esty{\widetilde{\theta}}^{(a)} - \esty{\theta}^{(a)} \\
  &= (\Pn - \PE) \esty{\phi}^{(a)}(Z; \mathcal{Q}^{\ast}) \ + \\
  &\qquad {(\Pn - \PE) \{\esty{\phi}^{(a)}(Z; \widehat{\mathcal{Q}}) - \esty{\phi}^{(a)}(Z; \mathcal{Q}^{\ast})\}} \ +\\
  &\qquad R(\widehat{\mathcal{Q}}),
\end{align*}
where the empirical process term,
$(\Pn - \PE) \{\esty{\phi}^{(a)}(Z; \widehat{\mathcal{Q}}) - \esty{\phi}^{(a)}(Z; \mathcal{Q}^{\ast})\}$,
can be controlled to be $o_{P}(n^{-1/2})$ even when the nuisance models for $Q$ and
$\rho$ are estimated with machine learning methods, as long as the nuisance
models and the corresponding influence function are learned using cross-fitting
\citep{chernozhukov_2018} and we assume that
$\widehat{Q}(X,a)$ and $Q^{\ast}(X,a)$ are bounded almost surely. For the
remainder term, we have
\begin{align*}
  R(\widehat{\mathcal{Q}}) =
                             &\underbrace{
  \esty{\widetilde{\theta}}^{(a)} - \esty{\theta}^{(a)} +
  \PE\left[  \frac{I(A=a)R}{\widehat{\pi}(a)\widehat{\rho}(a)}
                             (Y-\esty{\widetilde{\theta}}^{(a)})\right]
                             }_{\mathcal{S}_{1}}  \\
   + &\underbrace{
    \PE\left[
    \frac{\widehat{\pi(a)} - I(A=a)}{\widehat{\pi}(a)}\frac{\widehat{\rho}(X,a)}{\widehat{\rho}(a)}
    \{\widehat{Q}(X,a)-
                    \esty{\widetilde{\theta}}^{(a)}
                    \}
    \right]
                             }_{\mathcal{S}_{2}}
\end{align*}
and
\begin{align*}
\mathcal{S}_{1} = \frac{\pi(a)\rho(a) - \widehat{\pi}(a)\widehat{\rho}(a)}
{\widehat{\pi}(a)\widehat{\rho}(a)}\left(
  \esty{\widetilde{\theta}}^{(a)} - \esty{\theta}^{(a)}
  \right) = o_{P}(n^{-1/2})
\end{align*}
since $(\esty{\widetilde{\theta}}^{(a)} - \esty{\theta}^{(a)})=o_{P}(1)$ and
$(\pi(a)\rho(a) - \widehat{\pi}(a)\widehat{\rho}(a))(\widehat{\pi}(a)\widehat{\rho}(a))^{-1} = O_{P}(n^{-1/2})$.
Further,
\begin{align*}
  \mathcal{S}_{2} &= \frac{\{\widehat{\pi}(a) - \pi(a)\}}{\widehat{\pi}(a)}
                    \PE
                    \left[
                    \frac{\widehat{\rho}(X,a)}{\widehat{\rho}(a)}
                    \{\widehat{Q}(X,a)-
                    \esty{\widetilde{\theta}}^{(a)}
                    \}
                    \right] \\
  &=
\frac{1}{n}\sum_{i=1}^{n}\frac{I(A_i=a) - \pi(a)}{\pi(a)} \PE
    \left[
    \frac{\rho(X,a)}{\rho(a)}
    \{Q^{\ast}(X,a)-
    \esty{\theta}^{(a)}
    \}
  \right]
     \\
                  &+
                    \PE
                    \bigg[
\frac{\widehat{\rho}(X,a)}{\widehat{\pi}(a)\widehat{\rho}_{a}}
                    \{\widehat{Q}(X,a)-
                    \esty{\widetilde{\theta}}^{(a)}
                    \}
 -\\
                  &\qquad
\frac{\rho(X,a)}{\pi(a)\rho(a)}
                    \{Q^{\ast}(X,a)-
                    \esty{\theta}^{(a)}
                    \}
  \bigg]\{\widehat{\pi}(a)-\pi(a)\}.
\end{align*}
Note that $\pr \left\{ (\widehat{\pi}(a)-\pi(a))^2\right\}^{1/2} = O_P(n^{-1/2})$. Thus, the last term is $o_P(n^{-1/2})$ due to convergence and boundedness of the nuisance models and continuity. It follows that
\begin{align*}
  \sqrt{n}\{\esty{\widehat{\theta}}^{(a)} - \esty{\theta}^{(a)}\} &=
  \frac{1}{\sqrt{n}} \sum_{i=1}^{n} \esty{\phi}^{(a)}(Z_{i}; \mathcal{Q}^{\ast}) + \\
                                                      & \frac{1}{\sqrt{n}}\sum_{i=1}^{n} \frac{I(A_i=a)-\pi(a)}{\pi(a)}
                                                        \E\left[\frac{\rho^{\ast}(X,a)}{\rho(a)}\{Q^{\ast}(X,a)-\esty{\theta}^{(a)}\}\right]
                                                         +
  o_{P}(1) \\
  &= \frac{1}{\sqrt{n}}\sum_{i=1}^{n} \esty{\xi}^{(a)}(Z_{i}; \mathcal{Q}^{\ast}) + o_P(1),
\end{align*}
    and from the CLT that
\begin{align*}
  \sqrt{n}\{\esty{\widehat{\theta}}^{(a)} - \esty{\theta}^{(a)}\} \rightsquigarrow
  \mathcal{N}(0, \sigma^{2}),
\end{align*}
where the variance estimate \(\sigma^{2}\) can be consistently estimated from
the empirical variance of
\begin{align*}
  \esty{\xi}^{(a)}(Z; \widehat{\mathcal{Q}})
  =
  \esty{\phi}^{(a)}(Z; \widehat{\mathcal{Q}}) +
\frac{I(A=a)-\widehat{\pi}(a)}{\widehat \pi(a)}   \Pn\left[\frac{\widehat{\rho}(X,a)}{\widehat \rho(a)}\{\widehat{Q}(X,a)-\esty{\widetilde{\theta}}^{(a)}\}\right].
\end{align*}

\section{General case EIF and one-step estimation properties}
\label{sec:gen-eif}

Assume that we observe the full data $W = \{X, A, T^{\ast}, \IT{\ast}Y\}$.
The influence function for the initial plug-in estimator for $\esty{\theta}^{(a)}$ is given by
\begin{align}
  \label{eq:fullphi}
  \frac{I\{A = a\}}{\pi(a)}\frac{\IT{\ast}}{S(a)} \left \{Y - \esty{\theta}^{(a)}\right\},
\end{align}
Due to randomization, $A\indep X$, the complement to the nuisance tangent space is given by
\begin{align}\label{eq:compltangent-gen-eif}
  \mathcal{T}^{\perp} = \{ (A-\pi(1))h(X) \mid \E[h(X)^{2}] <\infty \},
\end{align}
for any measurable function $h$. The efficient influence function is now derived
as the projection onto the tangent space. Using the same calculations as in
Section \ref{sec:eif:thetay}, we get that
\begin{align}
  \esty{\phi}^{(a)}(W; P)
   &= \frac{I(A=a)}{\pi(a)}\frac{\IT{\ast}}{S(a)}\left\{Y - \esty{\theta}^{(a)}\right\} \\
  &\quad + \frac{\{\pi(a) - I(A = a)\}}{\pi(a)}\frac{S(X,a)}{S(a)}\left\{Q(X, a)-\esty{\theta}^{(a)}\right\}.
\end{align}
Now, consider the observed case
$Z =  \{X, A, T, \Delta(\tau), \IT{} R,  \IT{} R Y\}$.
The combined right-censoring and missing outcome process constitutes a monotone
coarsening mechanism associated with orthogonal nuisance tangent spaces
(including the randomization nuisance tangent space).
As described in \citep{tsiatis2006semiparametric, laan03:_unified_method_for_censor_longit}
the observed data efficient influence function is given by the following mapping
of the full data efficient influence function
\begin{align*}
\esty{\phi}^{(a)}(Z; P) = &\frac{\Delta(\tau)}{S_C(T \mid X,A)}\frac{R}{\rho(X,A)}\esty{\phi}^{(a)}(W; P)\\
  + &\int_{0}^\tau \frac{\E_P[ \esty{\phi}^{(a)}(W; P) \mid T \geq u, X, A]}{S_C(u-\mid X,A)}\,dM_{C}(u\mid X,A)\\
  - & \frac{\Delta(\tau)}{S_C(\tau|X,A)}I(T \geq \tau ) \frac{R-\rho(X,A)}{\rho(X,A)}  \E_P[ \esty{\phi}^{(a)}(W; P) \mid T \geq \tau, \Delta(\tau)=1, X, A].
\end{align*}
We note that for $u \leq \tau$, under the given assumptions, it holds that
\begin{align*}
\E\left[\IT{\ast}  |X, A, T\geq u \right] &= \frac{S(X, A)}{S(u|X, A)},
\end{align*}
and
\begin{align*}
\E\left[I(T^{\ast} \geq \tau) Y |T\geq u, X, A \right] &= \frac{S(X, A)}{S(u|X, A)}Q(X,A).
\end{align*}
Thus, we get that the efficient influence function is
\begin{align}
  &\esty{\phi}^{(a)}(Z; P) \nonumber \\
  =\,\,&\frac{I(A=a)}{\pi(a)}\frac{\Delta(\tau)}{S_C(\tau \mid X,A)} \frac{\IT{}}{S(a)} \frac{R}{\rho(X,A)} \left\{Y - \esty{\theta}^{(a)}(P)\right\} \label{eq:eif_esty_copy_term1} \\
  + &  \frac{I(A=a)}{\pi(a)}\frac{\Delta(\tau)}{S_C(\tau|X,A)} \frac{\IT{}}{S(a)} \frac{\rho(X,A)-R}{\rho(X,A)}\left\{Q(X,A) - \esty{\theta}^{(a)}(P)\right\}\label{eq:eif_esty_copy_term2} \\
  + &\frac{I(A=a)}{\pi(a)} \int_{0}^\tau \frac{1}{S(u|X, A)S_C(u- \mid X,A)}\,dM_{C}(u\mid X,A)\frac{S(X,a)}{S(a)}\left\{Q(X, a)-\esty{\theta}^{(a)}(P)\right\} \label{eq:eif_esty_copy_term3} \\
  + & \frac{\{\pi(a) - I(A = a)\}}{\pi(a)}\frac{S(X,a)}{S(a)}\left\{Q(X, a)-\esty{\theta}^{(a)}(P)\right\}. \label{eq:eif_esty_copy_term4}
\end{align}
By definition, the second order remainder is given by
\begin{align*}
R(P) = \PE \esty{\phi}^{(a)}(Z; P) + \esty{\widetilde{\theta}}^{(a)}(P) - \esty{\theta}^{(a)}(P_0)
\end{align*}
Firstly, the expectation of \eqref{eq:eif_esty_copy_term1} equals
\begin{align*}
 \PE\left(\frac{I(A=a)}{\pi(a)}\frac{\Delta(\tau)}{S_C(\tau \mid X,A)} \frac{\IT{}}{S(a)} \frac{\rho_0(X,a)}{\rho(X,A)} \left\{Q_0(X,A) - \esty{\theta}^{(a)}(P)\right\} \right).
\end{align*}
Secondly, the expectation of \eqref{eq:eif_esty_copy_term2} equals
\begin{align*}
  \PE\left (\frac{I(A=a)}{\pi(a)}\frac{\Delta(\tau)}{S_C(\tau|X,A)} \frac{\IT{}}{S(a)} \frac{\rho(X,A)-\rho_0(X,A)}{\rho(X,A)} \left\{Q(X, a)-\esty{\theta}^{(a)}(P)\right\} \right)
\end{align*}
Adding the above two expression together yields
\begin{align}
  &\PE\left (\frac{I(A=a)}{\pi(a)}\frac{\Delta(\tau)}{S_C(\tau|X,A)} \frac{\IT{}}{S(a)} \frac{\rho(X,A)-\rho_0(X,A)}{\rho(X,A)} \left\{Q(X,A) - Q_0(X,A) \right\} \right) \nonumber \\
  +& \PE\left(\frac{I(A=a)}{\pi(a)}\frac{\Delta(\tau)}{S_C(\tau \mid X,A)} \frac{\IT{}}{S(a)} \left\{Q_0(X,A) - \esty{\theta}^{(a)}(P) \right\}\right). \label{eq:eif_est_remainder_step1}
\end{align}
Next we note that \eqref{eq:eif_est_remainder_step1} can be written as
\begin{align}
  &\PE\left(\frac{I(A=a)}{\pi(a)}\frac{\Delta(\tau)}{S_C(\tau \mid X,A)} \frac{\IT{}}{S(a)} \left\{Q_0(X,A) - \esty{\theta}^{(a)}(P) \right\}\right) \nonumber\\
  =& \PE\left(\frac{I(A=a)}{\pi(a)}\frac{I(C \geq \tau)}{S_C(\tau \mid X,A)} \frac{\IT{\ast}}{S(a)} \left\{Q_0(X,A) - \esty{\theta}^{(a)}(P) \right\} \right)\nonumber \\
  =& \PE\left(\frac{I(A=a)}{\pi(a)}\frac{\IT{\ast}}{S(a)} \left\{Q_0(X,A) - \esty{\theta}^{(a)}(P) \right\}\right)\label{eq:eif_est_remainder_step3} \\
  -& \PE\left(\frac{I(A=a)}{\pi(a)}\int_0^\tau \frac{\IT{\ast}}{S_C(u-|X,A)} d M_C(u|X,A) \frac{1}{S(a)} \left\{Q_0(X,A) - \esty{\theta}^{(a)}(P) \right\}\right) \label{eq:eif_est_remainder_step2}
\end{align}
% (TODO: does the above also hold for  $\widehat P$?)
Adding \eqref{eq:eif_est_remainder_step2} and the expectation of
\eqref{eq:eif_esty_copy_term3} equals
\begin{align*}
  &\PE\left(\frac{I(A=a)}{\pi(a)} \int_{0}^\tau \frac{S(X,A)}{S(u|X, A)S_C(u-\mid X,A)}\,dM_{C}(u\mid X,A)\frac{1}{S(a)}\left\{Q(X, a) - \esty{\theta}^{(a)}(P) \right\}\right) \\
  - &\PE\left(\frac{I(A=a)}{\pi(a)}\int_0^\tau \frac{\IT{\ast}}{S_C(u-|X,A)} d M_C(u|X,A) \frac{1}{S(a)} \left\{Q_0(X,A) - \esty{\theta}^{(a)}(P) \right\}\right)\\
  = &\PE\left(\frac{I(A=a)}{\pi(a)} \int_{0}^\tau \frac{S(X,A)}{S(u|X, A)S_C(u-\mid X,A)}\,dM_{C}(u\mid X,A)\frac{1}{S(a)}\left\{Q(X, a) - \esty{\theta}^{(a)}(P) \right\}\right) \\
  - &\PE\left(\frac{I(A=a)}{\pi(a)}\int_0^\tau \frac{S_0(X,A)}{S_0(u|X,A)S_C(u-|X,A)} d M_C(u|X,A) \frac{1}{S(a)}  \left\{Q_0(X,A) - \esty{\theta}^{(a)}(P) \right\}\right)\\
  = & \PE\Bigg(\frac{I(A=a)}{\pi(a)}\frac{1}{S(a)}\int_0^\tau \Bigg(\frac{S(X,A)}{S(u-|X,A)} \left\{Q(X, a) - \esty{\theta}^{(a)}(P) \right\} -
  \\
  &\hspace*{10em} \frac{S_0(X,A)}{S_0(u|X,A)} \left\{Q_0(X, a) - \esty{\theta}^{(a)}(P) \right\}\Bigg) \frac{1}{S_C(u-|X,A)} d M_C(u|X,A)
    \Bigg)
\end{align*}
Next, adding \eqref{eq:eif_est_remainder_step3} and the expectation of
\eqref{eq:eif_esty_copy_term4} equals
\begin{align}
  &\PE\left(\frac{I(A=a)}{\pi(a)}\frac{\IT{\ast}}{S(a)} \left\{Q_0(X, a) - \esty{\theta}^{(a)}(P) \right\}\right)\\
  +&\PE\left\{\frac{\{\pi(a) - I(A = a)\}}{\pi(a)}\frac{S(X,a)}{S(a)}\left\{Q(X, a) - \esty{\theta}^{(a)}(P) \right\}\right\}\\
  =&\PE\Bigg\{\frac{\{\pi(a) - I(A = a)\}}{\pi(a)}\Bigg\{ \frac{S(X,A)}{S(A)}\left\{Q(X, a) - \esty{\theta}^{(a)}(P) \right\} - \\
  &\hspace*{10em}
  \frac{S_0(X,A)}{S(A)}\left\{Q_0(X, a) - \esty{\theta}^{(a)}(P) \right\} \ \Bigg\} \Bigg\}
  \\
  +&\PE\left(\frac{I(A = a)}{\pi(a)}\frac{I(T^\ast\geq \tau)}{S(a)} \left\{Q_0(X, a) - \esty{\theta}^{(a)}(P) \right\} \right). \label{eq:eif_est_remainder_step4}
\end{align}
Finally, adding $\esty{\theta}^{(a)}(P) - \esty{\theta}^{(a)}(P_0)$
and \eqref{eq:eif_est_remainder_step4} equals
\begin{align*}
  &\PE\left(\frac{I(A = a)}{\pi(a)}\frac{I(T^\ast\geq \tau)}{S(a)} \left\{Q_0(X, a) - \esty{\theta}^{(a)}(P) \right\}\right) + \esty{\theta}^{(a)}(P) - \esty{\theta}^{(a)}(P_0)\\
  =& \frac{\pi S(a) - \pi_0 S_0(a)}{\pi S(a)} \left\{\esty{\theta}^{(a)}(P) - \esty{\theta}^{(a)}(P_0)\right\}
\end{align*}
Combining the above results yields that the second order remainder equals
\begin{align*}
  &\PE\left (\frac{I(A=a)}{\pi(a)}\frac{\Delta(\tau)}{S_C(\tau|X,A)} \frac{\IT{}}{S(a)} \frac{\rho(X,A)-\rho_0(X,A)}{\rho(X,A)} \left\{Q(X,A) - Q_0(X,A) \right\} \right) \\
  + &\PE\Bigg(\frac{I(A=a)}{\pi(a)}\frac{1}{S(a)}\int_0^\tau \Bigg(\frac{S(X,A)}{S(u|X,A)} \left\{Q(X, a) - \esty{\theta}^{(a)}(P) \right\} -
      \\ &\hspace*{10em}
           \frac{S_0(X,A)}{S_0(u|X,A)} \left\{Q_0(X, a) - \esty{\theta}^{(a)}(P) \right\}\ \Bigg) \frac{1}{S_C(u-|X,A)} d M_C(u|X,A)\Bigg)\\
  + &\PE\left\{\frac{\{\pi(a) - I(A = a)\}}{\pi(a)}\left\{ \frac{S(X,A)}{S(A)}\left\{Q(X, a) - \esty{\theta}^{(a)}(P) \right\} - \frac{S_0(X,A)}{S(A)}\left\{Q_0(X, a) - \esty{\theta}^{(a)}(P) \right\} \ \right\} \right\}\\
  +&\frac{\pi S(a) - \pi_0 S_0(a)}{\pi S(a)} \left\{\esty{\theta}^{(a)}(P) - \esty{\theta}^{(a)}(P_0)\right\}
\end{align*}

%%%%%%%%%%%%%%%%%%%%%%%%%%%%%%%
%% General power considerations
%%%%%%%%%%%%%%%%%%%%%%%%%%%%%%%
\section{Some general power considerations}\label{sec:supplpower}
Here we give some further insights to the rejection regions of the proposed testing procedure for rejecting at least one of the hypotheses $H_{Y|T^{\ast}}$ and $H_{T^{\ast}}$ as well as for rejecting both hypotheses.  We next use these insights to argue that in scenarios with substantial positive correlation between the estimated target parameters our proposal will have higher disjunctive (win on at least one) power than the Bonferroni-Holm procedure under any alternative. Moreover we argue that our proposal will have higher conjunctive (win on all) power than the Bonferroni-Holm procedure in all correlation scenarios and under all alternatives. In the below derivations we fix $\alpha$ at 2.5\%. Consequently all derived thresholds and critical values are specific to this value. However, all derivations are easily repeated for any other choice of $\alpha$.   

We are going to view $SW_{n,H_{Y|T^{\ast}}\cap H_{T^{\ast}}}$ as a function of $z_{min}$ and $z_{max}$ for fixed $\rho$. For this purpose we use that that $SW_{n,H_{Y|T^\ast}\cap H_{T^\ast}}$ can be represented in terms of $z_{min}=\min\{z_{Y|T^{\ast}},z_{T^{\ast}}\}$ and $z_{max}=\max\{z_{Y|T^{\ast}},z_{T^{\ast}}\}$ as: 

\begin{eqnarray}\label{sw:repr}
 &&SW_{n,H_{Y|T^\ast}\cap H_{T^\ast}}=I\big(z_{max}\geq 0,\: z_{min}\leq\hat{\rho}\cdot z_{max}\big)\cdot z_{max}^{2}+\nonumber\\
 &&+I\big(z_{max}\geq 0,\: z_{min}\geq\hat{\rho}\cdot z_{max}\big)\frac{(z_{max}-z_{min})^{2}+2\cdot(1-\hat{\rho})\cdot z_{min}\cdot z_{max}}{1-\hat{\rho}^{2}}
\end{eqnarray}

\begin{figure}[htpb]
  \centering
  \includegraphics[width=0.7\textwidth]{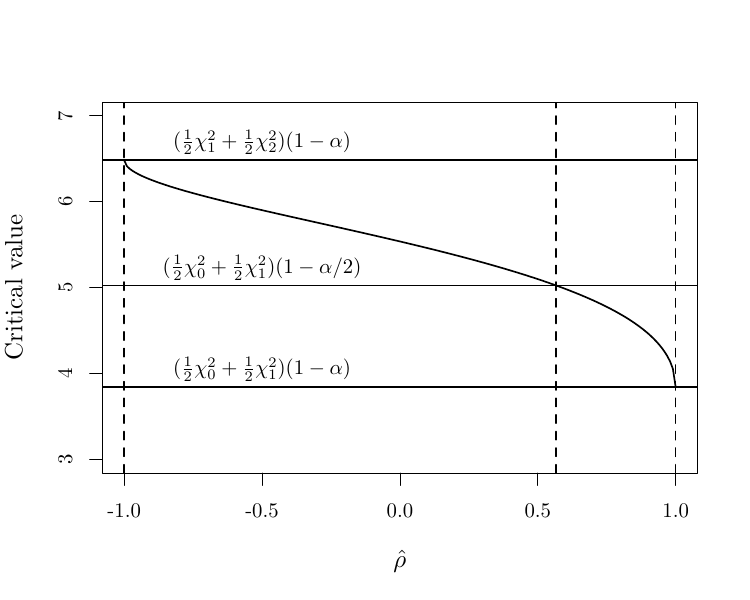}
  \caption{Critical values of $SW_{n,H_{Y|T^{\ast}}\cap H_{T^{\ast}}}$ as a function of correlation between estimators for $\alpha=0.025$. The solid lines mark the $1-\alpha$, $1-\alpha/2$, and $1-\alpha$ quantiles in the $\frac{1}{2}\chi^{2}_{1}+\frac{1}{2}\chi^{2}_{2}$, $\frac{1}{2}\chi^{2}_{0}+\frac{1}{2}\chi^{2}_{1}$, and $\frac{1}{2}\chi^{2}_{0}+\frac{1}{2}\chi^{2}_{1}$, respectively. Dashed lines mark the correlations where the critical values of $SW_{n,H_{Y|T^{\ast}}\cap H_{T^{\ast}}}$ equal these quantiles}
  \label{fig:critval}
\end{figure}

As a first step we evaluate the critical values of the intersection signed Wald test $SW_{n,H_{Y|T^{\ast}}\cap H_{T^{\ast}}}$ as a function of the estimated correlation $\hat{\rho}$ between the estimators. This can be done numerically by calculating $\hat{q}$ for each value of the correlation and then follow the steps described above with a fixed significance level $\alpha$. The resulting critical values are shown in Figure \ref{fig:critval}.

From a numerical search we find that for a correlation of $0.57$ the critical value of $SW_{n,H_{Y|T^{\ast}}\cap H_{T^{\ast}}}$ equals the $1-\alpha/2$ quantile in the $\frac{1}{2}\chi^{2}_{0}+\frac{1}{2}\chi^{2}_{1}$ distribution. We denote this quantile by $(\frac{1}{2}\chi^{2}_{0}+\frac{1}{2}\chi^{2}_{1})(1-\alpha/2)$ in what follows. Since the critical values of $SW_{n,H_{Y|T^{\ast}}\cap H_{T^{\ast}}}$ are decreasing as a function of correlation we note that for correlations above 0.57 the critical values of $SW_{n,H_{Y|T^{\ast}}\cap H_{T^{\ast}}}$ are below $(\frac{1}{2}\chi^{2}_{0}+\frac{1}{2}\chi^{2}_{1})(1-\alpha/2)$. 

In order to reject at least one of the hypotheses $H_{Y|T^{\ast}}$ or $H_{T^{\ast}}$ with the Bonferroni-Holm procedure it is required that   $I(z_{max}\geq0)z_{max}^{2}=\max\{SW_{n,H_{Y|T^{\ast}}},SW_{n,H_{T^{\ast}}}\}\ge(\frac{1}{2}\chi^{2}_{0}+\frac{1}{2}\chi^{2}_{1})(1-\alpha/2)$. It further follows from the representation (\ref{sw:repr}) and some straightforward calculations that $SW_{n,H_{Y|T^{\ast}}\cap H_{T^{\ast0}}}\geq I(z_{max}\geq0)z_{max}^{2}$. This means that for a correlation above $0.57$ we reject $SW_{n,H_{Y|T^{\ast}}\cap H_{T^{\ast}}}$ when we reject at least one hypothesis with the Bonferroni-Holm procedure. In this case we also reject at least one hypothesis with our proposal since $SW_{n,H_{Y|T^{\ast}}\cap H_{T^{\ast}}}$ is rejected and  $SW_{n,H_{Y|T^{\ast}}}$ or  $SW_{n,H_{T^{\ast}}}$ exceeds the $1-\alpha/2$  quantile  and therefore also the $1-\alpha$ quantile in the $\frac{1}{2}\chi^{2}_{0}+\frac{1}{2}\chi^{2}_{1}$ distribution.  

In summary, the above considerations show that for a correlation above $0.57$ a higher disjunctive power is ensured with our proposal compared to the Bonferroni-Holm procedure. 

Next, we turn to the conjunctive power, that is,  the probability of rejecting both hypotheses. We first note that in order for the Bonferroni-Holm procedure to reject both hypotheses it is required that $I(z_{max}\geq0)z_{max}^{2}>(\frac{1}{2}\chi^{2}_{0}+\frac{1}{2}\chi^{2}_{1})(1-\alpha/2)$ and  $I(z_{min}\geq0)z_{min}^{2}>(\frac{1}{2}\chi^{2}_{0}+\frac{1}{2}\chi^{2}_{1})(1-\alpha)$.

\begin{figure}[htpb]
  \centering
  \includegraphics[width=0.8\textwidth]{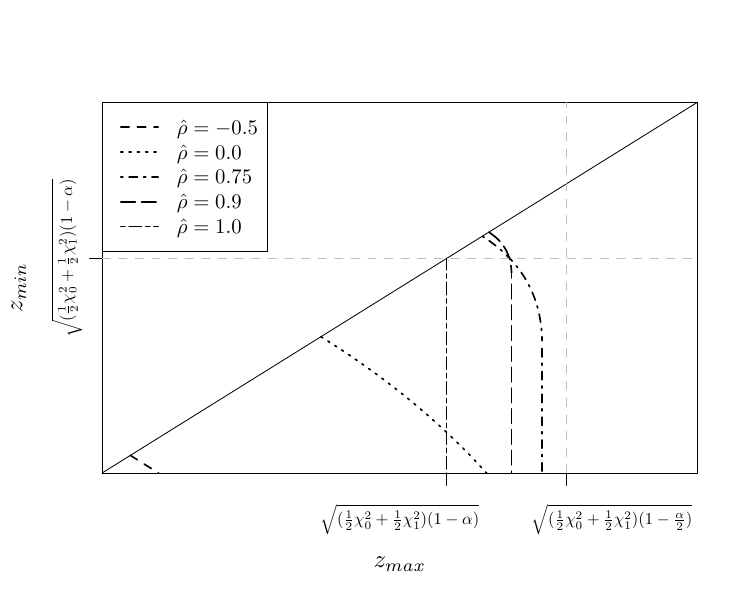}
  \caption{Dashed lines show the level curves of $SW_{n,H_{Y|T^{\ast}}\cap H_{T^{\ast}}}$ at the critical value ($\alpha=2.5\%$) for positive values of $z_{min}$ and $z_{max}$ and a range of correlations $\hat{\rho}$. }
  \label{fig:rejectreg}
\end{figure}

In Figure \ref{fig:rejectreg} we plotted the level curves of $SW_{n,H_{Y|T^{\ast}}\cap H_{T^{\ast}}}$ as a function of positive values of $z_{min}$ and $z_{max}$ for a range of fixed $\hat{\rho}$s. Since  $SW_{n,H_{Y|T^{\ast}}\cap H_{T^{\ast}}}$ is increasing on any line seqment it is clear from Figure \ref{fig:rejectreg} that any point in the conjunctive rejection region of the Bonferroni-Holm procedure is also rejected by the proposed procedure irrespective of the value of $\hat{\rho}$.

\begin{figure}[htpb]
  \centering
  \includegraphics[width=0.7\textwidth]{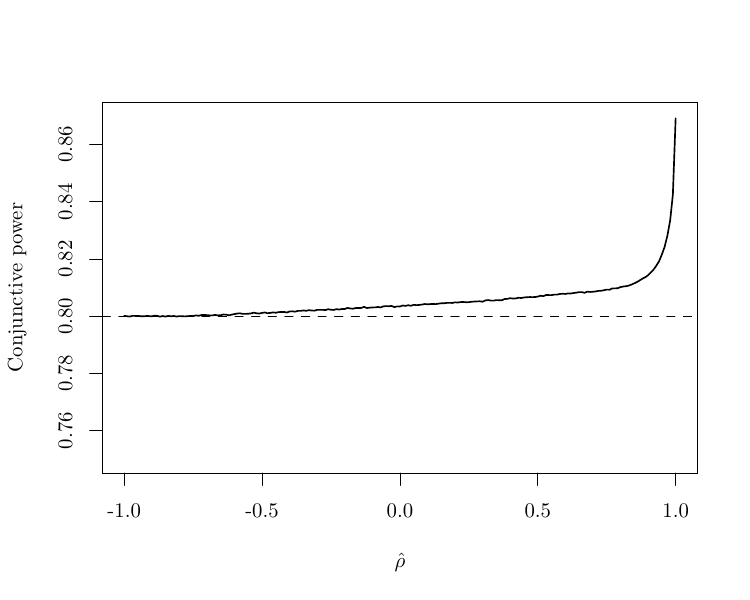}
  \caption{Conjunctive power as a function of $\hat{\rho}$. Solid line corresponds to the proposed testing procedure, dashed line corresponds to the Bonferroni-Holm procedure.}
  \label{fig:conjpower}
\end{figure}

To further gauge the actual power gain we calculate the conjunctive power of the proposed test strategy under a given alternative when testing using superiority/non-inferiority margins $\esty{\delta}=\estt{\delta}=0$ in $H_{T^{\ast}}$ and with $\alpha=0.025$. For each value of the correlation $\hat{\rho}$, the alternative is chosen to yield a non centrality parameter $(r(\hat{\rho}),r(\hat{\rho}))>0$  of $(z_{Y|T^{\ast}},z_{T^{\ast}})$ that will result in a conjunctive power of $80\%$ for the Bonferroni-Holm procedure. For each value of the correlation we calculate the conjunctive power of the proposed strategy by simulating 10 million realisations of $(z_{Y|T^{\ast}},z_{T^{\ast}})$ with the given non-centrality parameter and for each realization we then determine the outcome of the test strategy. Resulting conjunctive powers are plotted in Figure \ref{fig:conjpower} below as a function of the correlation.

\begin{figure}[htpb]
  \centering
  \includegraphics[width=0.7\textwidth]{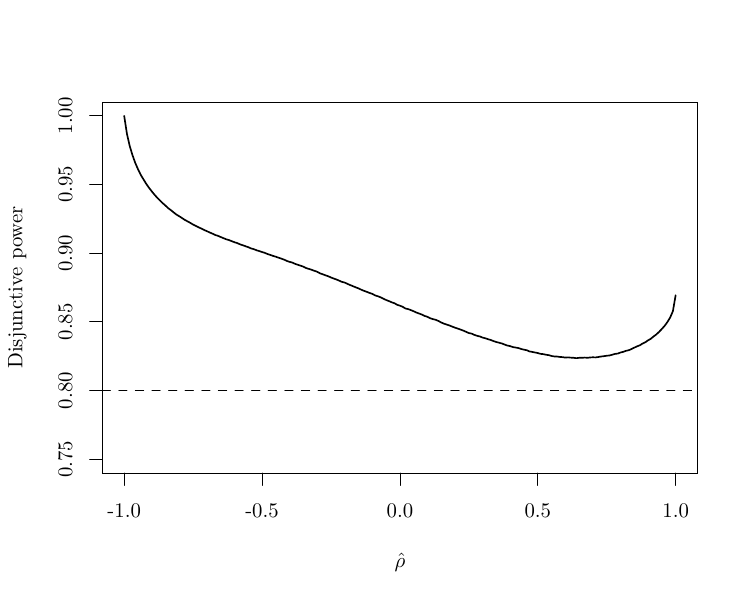}
  \caption{Disjunctive power as a function of $\hat{\rho}$. Solid line corresponds to the proposed testing procedure, dashed line corresponds to the Bonferroni-Holm procedure.}
  \label{fig:disjpower}
\end{figure}

Similarly we calculate the disjunctive power in a scenario with non-centrality parameter $$(r(\hat{\rho}),r(\hat{\rho}))>0,$$  chosen so that the disjunctive power of the Bonferroni-Holm procedure equals $80\%$. Resulting disjunctive powers are plotted in Figure \ref{fig:disjpower}.

%%%%%%%%%%%%%%%%%%%%%%%%%%%%%%%
%% R code example
%%%%%%%%%%%%%%%%%%%%%%%%%%%%%%%

\definecolor{fgcolor}{rgb}{0.345, 0.345, 0.345}
\newcommand{\hlnum}[1]{\textcolor[rgb]{0.686,0.059,0.569}{#1}}%
\newcommand{\hlsng}[1]{\textcolor[rgb]{0.192,0.494,0.8}{#1}}%
\newcommand{\hlcom}[1]{\textcolor[rgb]{0.678,0.584,0.686}{\textit{#1}}}%
\newcommand{\hlopt}[1]{\textcolor[rgb]{0,0,0}{#1}}%
\newcommand{\hldef}[1]{\textcolor[rgb]{0.345,0.345,0.345}{#1}}%
\newcommand{\hlkwa}[1]{\textcolor[rgb]{0.161,0.373,0.58}{\textbf{#1}}}%
\newcommand{\hlkwb}[1]{\textcolor[rgb]{0.69,0.353,0.396}{#1}}%
\newcommand{\hlkwc}[1]{\textcolor[rgb]{0.333,0.667,0.333}{#1}}%
\newcommand{\hlkwd}[1]{\textcolor[rgb]{0.737,0.353,0.396}{\textbf{#1}}}%
\let\hlipl\hlkwb

\makeatletter
\newenvironment{kframe}{%
 \def\at@end@of@kframe{}%
 \ifinner\ifhmode%
  \def\at@end@of@kframe{\end{minipage}}%
  \begin{minipage}{\columnwidth}%
 \fi\fi%
 \def\FrameCommand##1{\hskip\@totalleftmargin \hskip-\fboxsep
 \colorbox{shadecolor}{##1}\hskip-\fboxsep
     % There is no \\@totalrightmargin, so:
     \hskip-\linewidth \hskip-\@totalleftmargin \hskip\columnwidth}%
 \MakeFramed {\advance\hsize-\width
   \@totalleftmargin\z@ \linewidth\hsize
   \@setminipage}}%
 {\par\unskip\endMakeFramed%
 \at@end@of@kframe}
\makeatother
\definecolor{shadecolor}{rgb}{.97, .97, .97}
\definecolor{messagecolor}{rgb}{0, 0, 0}
\definecolor{warningcolor}{rgb}{1, 0, 1}
\definecolor{errorcolor}{rgb}{1, 0, 0}
\newenvironment{knitrout}{}{} % an empty environment to be redefined in TeX

\singlespacing

\section{Software implementation}\label{sec:software}

Installation of R package

\begin{Schunk}
\begin{Sinput}
> remotes::install_github("kkholst/targeted", ref="trunc-score")
\end{Sinput}
\end{Schunk}

\subsection{Simulation setup}
\begin{Schunk}
\begin{Sinput}
> ## Treatment assignment
> p.a <- 0.5
> ## SGLT2 at baseline
> p.x2 <- 0.156
> ## eGFR at baseline
> m.x1 <- list("x2=0" = 46.24, "x2=1" = 51.15)
> s.x1 <- list("x2=0" = 14.99, "x2=1" = 15.33)
> ## eGFR at landmark
> b.y <- list(
+   "a=0" = c(40.141, 0.895, 1.993),
+   "a=1" = c(43.121, 0.863, 2.620)
+ )
> s.y <- list("a=0" = 11.85, "a=1" = 12.16)
> ## Censoring
> b.e0 <- list(
+   "a=0" = c(log(0.00014), 0, 0),
+   "a=1" = c(log(9.35e-5), 0, 0)
+ )
> gamma.e0 <- list("a=0" = 6.691, "a=1" = 6.946)
> ## Primary event
> b.e1 <- list(
+   "a=0" = c(log(0.0285), -0.0243, -0.5832),
+   "a=1" = c(log(.01817), -0.0289, -0.1261)
+ )
> gamma.e1 <- list("a=0" = 1.822, "a=1" = 1.901)
> ## Death other causes
> b.e2 <- list(
+   "a=0" = c(log(0.0154), -0.0205, -0.4549),
+   "a=1" = c(log(0.0160), 0.00687, -0.598)
+ )
> gamma.e2 <- list("a=0" = 1.143, "a=1" = 1.071)
> ## Missing data mechanism
> b.r <- list(
+   "a=0" = c(2.243, 0, 0),
+   "a=1" = c(2.309, 0, 0)
+ )
> pars <- list(
+   a = p.a,
+   x1 = list(m = m.x1, sd = s.x1),
+   x2 = p.x2,
+   y = list(m = b.y, sd = s.y),
+   r = b.r,
+   t0 = list(m = b.e0, shape = gamma.e0),
+   t1 = list(m = b.e1, shape = gamma.e1),
+   t2 = list(m = b.e2, shape = gamma.e2)
+ )
> 
> 
> simdata <- function(n, # sample-size
+                     parameters = pars, # model parameter
+                     tau = 2, # landmark time
+                     null = FALSE
+                     ) {
+   a <- rbinom(n, 1, parameters[["a"]]) # treatment variable
+   x2 <- rbinom(n, 1, parameters[["x2"]]) # SGL2 treatment at baseline
+   x1 <- rnorm(n, # eGFR at baseline
+     mean = with(parameters[["x1"]], m[["x2=0"]] * (1 - x2) + m[["x2=1"]] * x2),
+     sd = with(parameters[["x1"]], sd[["x2=0"]] * (1 - x2) + sd[["x2=1"]] * x2)
+   )
+   mean.x1 <- with(parameters[["x1"]], m[["x2=0"]] * (1 - parameters[["x2"]]) +
+                                       m[["x2=1"]] * parameters[["x2"]])
+ 
+   placebo <- "a=0"
+   active <- ifelse(null, "a=0", "a=1")
+   # Design matrix
+   X <- cbind(1, x1 - mean.x1, x2)
+   # Latent clinical outcome (eGFR)
+   y0 <- rnorm(n,
+     mean = with(parameters[["y"]], X %*% m[[placebo]] * (1 - a) +
+       X %*% m[[active]] * a),
+     sd = with(parameters[["y"]], sd[[placebo]] * (1 - a) + sd[[active]] * a)
+   )
+   sim_weibull <- function(X, a, gamma, b) {
+     shape <- gamma[[placebo]] * (1 - a) + gamma[[active]] * a
+     lp <- X %*% b[[placebo]] * (1 - a) + X %*% b[[active]] * a
+     rweibull(n, shape = shape, scale = exp(lp / -shape))
+   }
+   # latent censoring time
+   t0 <- sim_weibull(X, a, parameters[["t0"]]$shape, parameters[["t0"]]$m)
+   # latent event time
+   t1 <- sim_weibull(X, a, parameters[["t1"]]$shape, parameters[["t1"]]$m)
+   # latent competing death event time
+   t2 <- sim_weibull(X, a, parameters[["t2"]]$shape, parameters[["t2"]]$m)
+   failure.time <- pmin(t1, t2)
+   time <- pmin(t0, t1, t2)
+   status <- apply(cbind(t0, t1, t2), 1, which.min) - 1
+   # Observation indicator given T>tau
+   p.r <- lava::expit(X %*% parameters[["r"]][[placebo]] * (1 - a) +
+     X %*% parameters[["r"]][[active]] * a)
+   r.tau <- rbinom(n, 1, p.r)
+   y0[failure.time < tau] <- NA
+   # Observed clinical outcome (eGFR)
+   y <- y0
+   y[r.tau == 0 & time < tau] <- NA
+   # Return combined data
+   d <- data.frame(a, x1, x2, y0, y, time, r0=r.tau,
+     r = (!is.na(y)) * 1, status, failure.time
+   )
+   return(d)
+ }
\end{Sinput}
\end{Schunk}

\subsection{Estimation procedure}

\begin{Schunk}
\begin{Sinput}
> set.seed(1)
> dat <- simdata(n = 4000)
> head(dat)
\end{Sinput}
\begin{Soutput}
  a       x1 x2         y0          y     time r0 r status failure.time
1 0 29.23189  0 27.3580958 27.3580958 3.160429  1 1      1     3.160429
2 0 57.70071  0 39.8837977 39.8837977 3.472898  1 1      0     8.671925
3 1 54.79494  0 32.0694152 32.0694152 2.667604  1 1      0     3.701509
4 1 25.97811  0 45.4512843 45.4512843 3.458985  1 1      1     3.458985
5 0 20.03186  1 -0.3080661 -0.3080661 2.431071  1 1      0     4.619181
6 1 55.09128  0 47.7774394 47.7774394 3.943634  1 1      0     5.538106
\end{Soutput}
\end{Schunk}

\begin{Schunk}
\begin{Sinput}
> mod1 <- learner_glm(y ~ a * (x1 + x2))
> mod2 <- learner_glm(r ~ a * (x1 + x2), family = binomial)
> est <- estimate_truncatedscore(
+   data = dat,
+   mod.y = mod1,
+   mod.r = mod2,
+   mod.a = a ~ 1,
+   mod.event = mets::Event(time, status>0) ~ a * (x1+x2),
+   time = 2, 
+   cens.code = 0,
+ )
> 
> est
\end{Sinput}
\begin{Soutput}
               Estimate  Std.Err     2.5%    97.5%   P-value
E(Y|T>2.0,A=0) 40.52349 0.368697 39.80085 41.24612 0.000e+00
E(Y|T>2.0,A=1) 43.65669 0.364699 42.94190 44.37149 0.000e+00
diff            3.13321 0.425018  2.30019  3.96623 1.682e-13
--------------                                              
P(T>2.0|A=0)    0.87824 0.007197  0.86413  0.89234 0.000e+00
P(T>2.0|A=1)    0.90832 0.006544  0.89550  0.92115 0.000e+00
riskdiff        0.03009 0.009710  0.01105  0.04912 1.946e-03
\end{Soutput}
\end{Schunk}

\begin{Schunk}
\begin{Sinput}
> s <- summary(est, noninf.y = 0, noninf.t = -0.05)
> s
\end{Sinput}

\begin{Soutput}
-- Parameter estimates --
\end{Soutput}

\begin{Soutput}
               Estimate  Std.Err     2.5%    97.5%   P-value
E(Y|T>2.0,A=0) 40.52349 0.368697 39.80085 41.24612 0.000e+00
E(Y|T>2.0,A=1) 43.65669 0.364699 42.94190 44.37149 0.000e+00
diff            3.13321 0.425018  2.30019  3.96623 1.682e-13
--------------                                              
P(T>2.0|A=0)    0.87824 0.007197  0.86413  0.89234 0.000e+00
P(T>2.0|A=1)    0.90832 0.006544  0.89550  0.92115 0.000e+00
riskdiff        0.03009 0.009710  0.01105  0.04912 1.946e-03
\end{Soutput}
\begin{Soutput}
-- One-sided tests --
\end{Soutput}
\begin{Soutput}

b1 = E(Y|T>2.0,A=1) - E(Y|T>2.0,A=0)

	Signed Wald Test

data:  H0: b1 =< 0
Q = 54.346, p-value = 8.408e-14
alternative hypothesis: HA: b1 > 0
sample estimates:
      b1 
3.133207 


b2 = P(T>2.0|A=1) - P(T>2.0|A=0)

	Signed Wald Test

data:  H0: b2 =< -0.05
Q = 68.022, p-value < 2.2e-16
alternative hypothesis: HA: b2 > -0.05
sample estimates:
        b2 
0.03008579 
\end{Soutput}
\begin{Soutput}
-- Intersection test --
\end{Soutput}
\begin{Soutput}

	Signed Wald Intersection Test

data:  
Intersection null hypothesis: b =< [0, -0.05]
w = [0.5, 0.5]
Q = 33.461, p-value < 2.2e-16
\end{Soutput}
\end{Schunk}

Extracting the test statistics and p-values in a matrix-form
\begin{Schunk}
\begin{Sinput}
> parameter(s)
\end{Sinput}
\begin{Soutput}
               estimate statistic      p.value
b1           3.13320718  54.34558 8.407769e-14
b2           0.03008579  68.02157 8.085827e-17
intersection         NA  33.46054 0.000000e+00
\end{Soutput}
\end{Schunk}

%%%%%%%%%%%%%%%%%%%%%%%%%%%%%%%

\end{document}